\documentclass[conference]{IEEEtran}

\IEEEoverridecommandlockouts
\usepackage{cite}
\usepackage{amsmath,amssymb,amsfonts}
\usepackage{graphicx}
\usepackage{textcomp}
\usepackage{xcolor}
\usepackage{bbding}
\usepackage{multirow}
\usepackage{caption}
\usepackage{subcaption}
\usepackage[normalem]{ulem}
\usepackage{authblk}
\usepackage{algorithm}
\usepackage{algpseudocode}

\def\BibTeX{{\rm B\kern-.05em{\sc i\kern-.025em b}\kern-.08em
    T\kern-.1667em\lower.7ex\hbox{E}\kern-.125emX}}
\begin{document}

\title{Profile-Guided Parallel Task Extraction and Execution for Domain Specific Heterogeneous SoC \\
\thanks{
% Place holder for acknowledgment. 
This material is based on research sponsored by Air Force Research Laboratory (AFRL) and Defense Advanced Research Projects Agency (DARPA) under agreement number FA8650-18-2-7860. The U.S. Government is authorized to reproduce and distribute reprints for Governmental purposes notwithstanding any copyright notation thereon. The views and conclusion contained herein are those of the authors and should not be interpreted as necessarily representing the official policies or endorsements, either expressed or implied, of Air Force Research Laboratory (AFRL) and Defence Advanced Research Projects Agency (DARPA) or the U.S. Government.
}
}

\author[1]{Liangliang Chang}
\author[2]{Joshua Mack}
\author[1]{Benjamin Willis}
\author[1]{Xing Chen}
\author[1]{John Brunhaver}
\author[2]{\\Ali Akoglu}
\author[1]{Chaitali Chakrabarti}
\affil[1]{The School of Electrical, Computer and Energy Engineering, Arizona State University, USA}
\affil[2]{Electrical and Computer Engineering Department, The University of Arizona, USA}
\affil[ ]{\textit{\{lchang21,bwilli46,xchen382,jbrunhav,chaitali\}@asu.edu}, \textit{\{jmack2545,akoglu\}@arizona.edu}}

% \author{\IEEEauthorblockN{Liangliang Chang}
% \IEEEauthorblockA{\textit{Arizona State University, USA} \\
% \and
% \and
% \and
% \and
% \and

\maketitle

\begin{abstract}
% \josh{Across all experiments, we observe an up-to X\% speedup versus a baseline statically-accelerated execution flow...}
%An End to End Automatic System Speeds Up Domain-Specific SoC By Leveraging Task-Level Parallelism Based on Dynamic Analysis
In this study, we introduce a methodology for automatically transforming user applications in the radar and communication domain written in C/C++ based on dynamic profiling to a parallel representation targeted for a heterogeneous SoC.  We present our approach for instrumenting the user application binary during the compilation process with barrier synchronization primitives that enable runtime system schedule and execute independent tasks concurrently over the available compute resources. We demonstrate the capabilities of our integrated compile time and runtime flow through task-level parallel and functionally correct execution of real-life applications. We perform validation of our integrated system by executing four distinct applications each carrying various degrees of task level parallelism  over the Xeon-based multi-core homogeneous processor. We use the proposed compilation and code transformation methodology to re-target each application for execution on a heterogeneous SoC composed of three ARM cores and one FFT accelerator that is emulated on the Xilinx Zynq UltraScale+ platform. We demonstrate our runtime's ability to process application binary, dispatch independent tasks over the available compute resources of the emulated SoC on the Zynq FPGA based on three different scheduling heuristics. Finally we demonstrate execution of each application individually with task level parallelism on the Zynq FPGA and execution of workload scenarios composed of multiple instances of the same application as well as mixture of two distinct applications to demonstrate ability to realize both application and task level parallel execution.  
%\liang{We demonstrate the capabilities of our integrated compile time and runtime flow on multiple software and hardware configurations. We evaluate the performance of single application execution and dynamically arriving application execution scenarios on three different computation platforms including an X86-based multicore processor, a simulation-based SoC, and a real SoC emulated using a Xilinx Zynq UltraScale+ platform.}
Our integrated approach offers a path forward for application developers to take full advantage of the target SoC without requiring users to become hardware and parallel programming experts.
\end{abstract}


\section{Introduction}
%\subsection{Prior Works}
SoCs composed of a pool of heterogeneous processing elements offer performance gains over their homogeneous counterparts as they allow pairing each task or execution phase of application with a suitable processing element (PE) based on the state of the system resources.  
To harness this flexibility, programming models have been introduced where application developers or domain experts guide the compilation process by making task to PE mapping decisions based on offline profiling. 
For example, in CUDA-based programming, programmers have to understand the application, partition it into independent tasks, and manually map them to threads and blocks in GPU.
This model of computation results in a static execution flow and a hand-crafted schedule that is greedily tuned for a single application. 
%\sout{When we consider non-uniform performance across the processing elements, such static application deployment would result with tasks starve for compute resources in scenarios where multiple dynamically arriving applications share the system.} 
In this study, we introduce an integrated compile time and runtime environment that automatically detects parallelism in the user application, transforms the program to parallel representation, and provides the runtime system with a flexible binary structure. This allows the runtime system to dynamically schedule and launch these tasks in parallel to heterogeneous resources based on the system state rather than rely on a hand-crafted static schedule.

\begin{table}[!t]
\centering
\scalebox{0.73}{
\begin{tabular}{|l|c|cc|cc|c|c|}
\hline
\multirow{2}{*}{\textbf{}} &
  \multicolumn{1}{c|}{\multirow{2}{*}{\textbf{\begin{tabular}[c]{@{}c@{}} Dedicated\\ Framework\end{tabular}}}} &
  \multicolumn{2}{c|}{\textbf{Program Analysis}} &
  \multicolumn{2}{c|}{\textbf{Target Architecture}} &
  \multirow{2}{*}{\textbf{End-to-End}} \\ \cline{3-6}
  % \multicolumn{1}{c|}{\multirow{2}{*}{\textbf{\begin{tabular}[c]{@{}c@{}}Compile Time\\ Task Offload\end{tabular}}}} \\ \cline{3-6}
  \multicolumn{1}{|c|}{} &
   &
  \multicolumn{1}{c|}{\textbf{Static}} &
  \textbf{Dynamic} &
  \multicolumn{1}{c|}{\textbf{Multicore}} &
  \textbf{Hetero. SoC} &
   \\ \hline
  Tensorflow\cite{abadi2016tensorflow} & \CheckmarkBold & \multicolumn{1}{c|}{\CheckmarkBold} &   & \multicolumn{1}{c|}{\CheckmarkBold} & \CheckmarkBold & \CheckmarkBold   \\ \hline
Halide\cite{ragan2013halide}     & \CheckmarkBold & \multicolumn{1}{c|}{\CheckmarkBold} &   & \multicolumn{1}{c|}{\CheckmarkBold} &   & \CheckmarkBold  \\ \hline
HPVM\cite{kotsifakou2018hpvm}       & \CheckmarkBold & \multicolumn{1}{c|}{\CheckmarkBold} &   & \multicolumn{1}{c|}{\CheckmarkBold}  & \CheckmarkBold & \CheckmarkBold    \\ \hline
Chi et al.\cite{chi2021extending}      & \CheckmarkBold & \multicolumn{1}{c|}{\CheckmarkBold} &   & \multicolumn{1}{c|}{}  & \CheckmarkBold & \CheckmarkBold    \\ \hline
Parwiz \cite{ketterlin2012profiling}      &   & \multicolumn{1}{c|}{\CheckmarkBold} & \CheckmarkBold & \multicolumn{1}{c|}{\CheckmarkBold} &   &         \\ \hline
SD3 \cite{kim2010sd3}        &   & \multicolumn{1}{c|}{\CheckmarkBold}  & \CheckmarkBold & \multicolumn{1}{c|}{\CheckmarkBold} &   &         \\ \hline
Wang et al. \cite{wang2014integrating}       &   & \multicolumn{1}{c|}{\CheckmarkBold}  & \CheckmarkBold & \multicolumn{1}{c|}{\CheckmarkBold} &   & \CheckmarkBold \\ \hline
Kremlin \cite{garcia2011kremlin} &   & \multicolumn{1}{c|}{\CheckmarkBold} & \CheckmarkBold & \multicolumn{1}{c|}{\CheckmarkBold}  &  & \CheckmarkBold   \\ \hline
Ours       &   & \multicolumn{1}{c|}{\CheckmarkBold} & \CheckmarkBold & \multicolumn{1}{c|}{\CheckmarkBold}  & \CheckmarkBold & \CheckmarkBold \\ \hline
\end{tabular}
}
\caption{Related work on program analysis and workload parallelization}
\vspace{-6mm}
\label{tab:related-work}
\end{table}

Programming models such as OpenMP\cite{dagum1998openmp} and Pthread\cite{butenhof1997programming} offer interfaces like pragma labels and thread bindings where users specify the parallelism in their applications explicitly. 
Program analysis tools have been introduced to enable transforming a user application to a parallel representation based on static ~\cite{ragan2013halide, abadi2016tensorflow, chi2021extending, kotsifakou2018hpvm} or dynamic \cite{kim2010sd3,garcia2011kremlin,wang2014integrating,ketterlin2012profiling} analysis as listed in Table~\ref{tab:related-work}. 
These tools also vary in terms of their approach to parallelism detection from instruction level granularity targeted for multi-core architectures to task level granularity for heterogeneous architectures. 
%\liang{Program parallelization has been studied for decades; popular solutions still rely on the programmer to have the application's computation flow in mind and write parallel applications manually. Existing works like OpenMP\cite{dagum1998openmp} and Pthread\cite{butenhof1997programming} support parallel code generation based on interfaces like pragma labels and thread bindings. Others support task-level parallel offloading based on program analysis to help the application get better performance on heterogeneous or multicore architectures. } 
%Table~\ref{tab:related-work} shows the existing works on workload parallelization based on program analysis.
Static parallelization methods require the user to express the application in an explicit data flow style either using a domain-specific language (DSL) or a dedicated framework. 
%\sout{It reduces the complexity of program analysis but requires extra effort from framework developers and application programmers. Also, static data flow analysis is difficult in languages like C and C++ that provide a rich set of control flow structures, including pointers.} 
Among the dynamic methods, our approach is the only one that targets both homogeneous multi-core architectures and heterogeneous SoCs in an end-to-end integrated compile and run time flow.
%\sout{Dynamic methods can resolve many of those cases, but existing methods typically target multi-core architectures. These methods perform data dependence analysis within a function call or loop.} 
%short instruction distance and are limited by the local variable scope of control structures like functions and loops. 
%\sout{Therefore, the code generation techniques of existing profile-guided works\ali{, while performing well for multi-core architectures,} are not suitable for transforming the program to support coarse-grained task-level parallelism. }

%\liang{Parwiz\cite{ketterlin2012profiling} and SD3\cite{kim2010sd3} produce parallel recommendations by commenting on the source code, while in terms of end-to-end system integration, others\cite{abadi2016tensorflow,ragan2013halide,chi2021extending,wang2014integrating,garcia2011kremlin} support parallel code generation. They have optimized flow to map the task to suitable hardware resources at compile time considering locality or bandwidth. For the works targeting heterogeneous SoC, HPVM\cite{kotsifakou2018hpvm} provides support for runtime scheduling. Our work supports parallel code generation and compiles the code of tasks to access hardware. Our runtime can dynamically map the task to suitable hardware based on availability and schedule policy.}

%Unlike static and dynamic analysis methods, 
Our approach employs a profile-guided dynamic analysis of memory access patterns using a combination of memory tuples, loop-access patterns and function pointers for inferring the required runtime control states that are necessary for task-level dependence analysis. It offers the ability to retarget user applications with coarse-grained computation tasks for heterogeneous architectures through code transformations that enable parallel task execution in one unified compile time and runtime framework. The key technical contributions of this study are as follows:

\begin{itemize}
\item We introduce a novel profile-guided memory analysis approach to detect the data dependencies among coarse-grained tasks in a given application and expose the parallelism among those tasks.
\item We present a methodology that partitions the user application into serial and parallel tasks following a fork-join programming model and compiles into an application binary representation with embedded parallelism and instrumentation such that the runtime system can issue and execute all independent tasks concurrently.
%\item \sout{We introduce a methodology to refactor and generate the new application representation such that our runtime system can issue and execute all parallel tasks. }
\item We integrate the profile-guided parallel program generation tool flow with an open-source runtime and demonstrate an end-to-end system that is able to compile and execute real-life applications on off-the-shelf platforms.
\end{itemize}

%Our work addresses mapping tasks represented by coarse-grained computation kernels on heterogeneous SoC. Instead of dedicated programming interfaces like DSL or frameworks, we only need the programmer to use a lightweight computation library providing the task APIs. 
%We utilize profile-guided analysis for inferring the required runtime control states that are necessary for task-level dependence analysis. 
%The memory analysis complexity is reduced by extending point-based to a tuple-based algorithm and performing preprocessing transformations to assist the analysis.

We demonstrate the ability to identify task-level parallelism for real life applications, transform user-application for parallel execution, and successfully execute those tasks in parallel first using an event-driven simulation environment DS3\cite{arda2020ds3}. After validating the ability to extract parallelism, we demonstrate functionally correct task-level parallelism in our runtime for those parallelized applications on a homogeneous 8-core Xeon processor. Finally, we demonstrate functionally correct parallel execution on an emulated heterogeneous SoC composed of 3 ARM CPUs and an FPGA-based accelerator. This emulation platform illustrates not only our ability to process single application with parallel execution flow but also our ability to execute multiple dynamically arriving applications supporting both task-level and application level parallel execution across heterogeneous set of resources.

% \liang{For the real work radar applications, considering the computation parts, on the X86-based multicore processor platform, we achieve an average of 44\% of time-saving. 
% On the simulation-based SoC with one processor and multiple FFT/GEMM accelerators, the average time saving is up to 75.6\%.}
% On the FPGA-based emulation platform, which is close to the real scenario, we perform evaluations involving two applications and three scheduling heuristics, and we find that across our heaviest workload, we achieve a 1.5x speedup over a baseline serial implementation that greedily favors accelerator execution.

\begin{figure*}[!h]
\centerline{\includegraphics[width=0.98\textwidth]{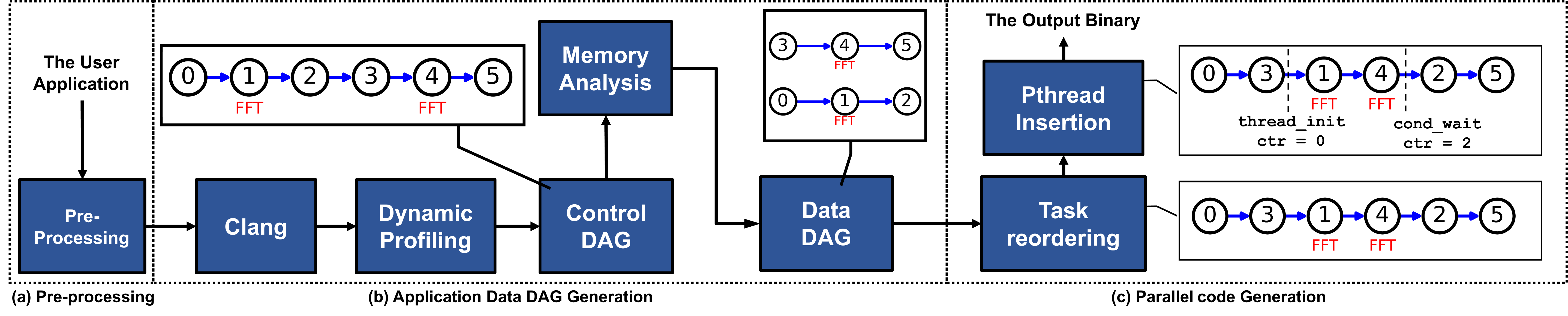}}
\caption{Overall tool flow involves \emph{Pre-processing} (Section~\ref{sec:preprocess}). \emph{Data DAG Generation} (Section~\ref{sec:data-dag}), followed by \emph{Parallel Code Generation} (Section~\ref{sec:codegen}).}
% \vspace{-4mm}
\label{fig:ToolFlow}
\end{figure*}

\begin{figure*}[!ht]
\centerline{\includegraphics[width=1.0\textwidth]{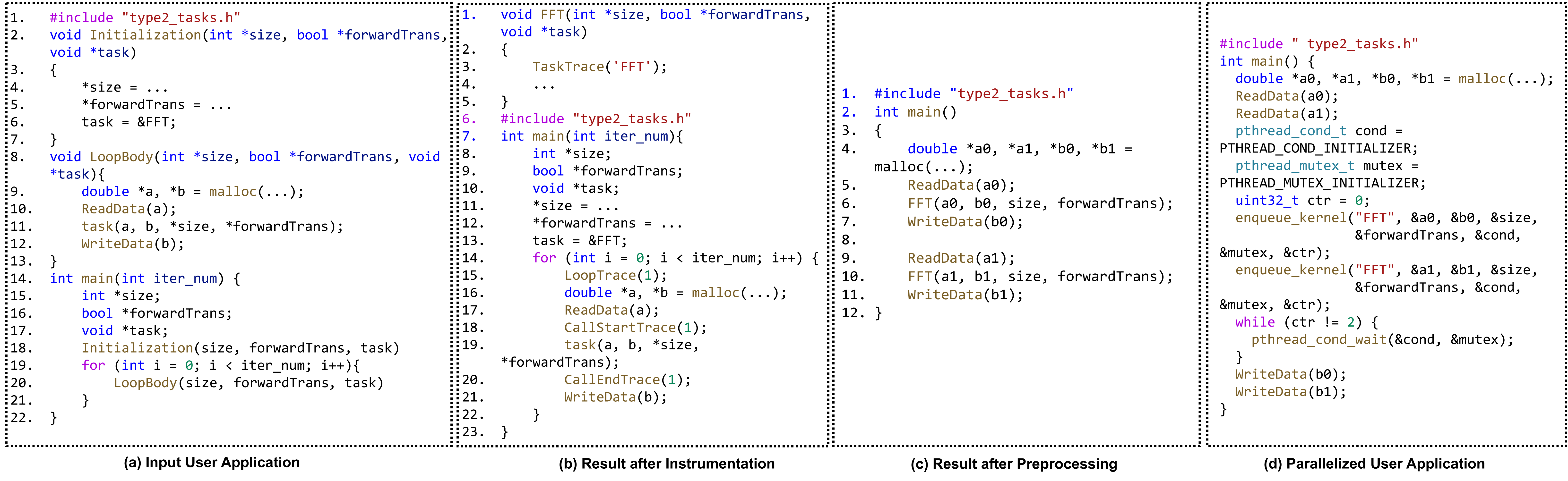}}
\caption{The state of the user application as it is processed through based on the tool flow shown in Figure~\ref{fig:ToolFlow}. (a) User application with function pointers, pointers for passing variables and unknown loop iteration count; (b) Instrumented user application for tracing during pre-processing; (c) Flattened user application after pre-processing step; (d) Parallelized user application instrumented with counter based conditional wait.}
\vspace{-4mm}
\label{fig:code-example}
\end{figure*}

% \begin{figure}[!ht]
% \centerline{\includegraphics[width=0.95\columnwidth]{updated-figure/preprocessing.pdf}}
% \caption{
% \liang{\emph{Pre-processing} of the user application.
% In each step, \emph{Special Call Instructions} are instructions that call a task API or a function with an unknown module. \emph{Special Loops} are loops that have \emph{Special Call Instructions} in the loop body. \emph{Special Functions} are functions that have \emph{Special Call Instructions} in the function body.}
% }
% \vspace{-4mm}
% \label{fig:preprocessing}
% \end{figure}

\section{Profile Guided Parallel Program Generation}
\label{sec:methodology}
\subsection{Overview}\label{sec:overview}
%In this section, we introduce the details on how we generate the parallel program step by step. We provide an overview of our tool flow and an example application that will be used to help understand the different steps. 

% Figure~\ref{fig:preprocessing} shows a preprocessing flow that helps the following analysis and transformation.\liang{In Figure~\ref{fig:preprocessing}-\emph{label 1}, we show a typical radar application.}  It will be transformed into 
% as illustrated in Figure~\ref{fig:ToolFlow}-\emph{label 4}. 

The overall flow of the profile guided parallel program generation as illustrated in Figure~\ref{fig:ToolFlow}, involves the \textit{Preprocessing} (Step-1) and \textit{Application Data DAG Generation} (Step-2) followed by \textit{Parallel Code Generation} (Step-3).
In this design flow, we leverage TraceAtlas~\cite{uhrie2020automated} for profile-based program analysis and update Compiler Integrated Extensible DSSoC Runtime (CEDR)~\cite{MackTECS22} for runtime task scheduling. 
%In the following paragraphs we present the details of our approach starting with an overview of the explain this process with the user written code in C/C++ as a running example.
%\subsubsection{Dynamic Tracing}
TraceAtlas\cite{uhrie2020automated} offers flexible interfaces to support dynamic profiling and trace analysis of LLVM IR~\cite{lattner2004llvm}  
%\sout{provides flexible interfaces for  dynamic tracing and trace analysis}
rapidly with lower resource requirements compared to other frameworks~\cite{aladdinShao2014,Wasabi2019Lehmann,luk2005pin}. 
%\sout{The overheads of long tracing time, long analyzing time, and high storage consumption are well addressed in TraceAtlas. The disk storage is optimized by compressing traced data using Zlib. In order to speed up, the data is collected to an optimal size before it is compressed.} 
%\sout{TraceAtlas is built on top of the LLVM framework, it supports the tracing of LLVM IR.} 
We use TraceAtlas 
%coupled with our memory analysis algorithms 
to collect memory address ranges accessed by each basic block in the IR along with the runtime control states of those blocks. 
We then use this information to identify tasks that can be executed in parallel during a task-level memory analysis of the user application. 
%After the source code is compiled, we can trace the needed IR instructions and corresponding operands by specifying them using the data writing function in TraceAtlas. 
%Instead of writing the whole IR, the user can encode the target data, which will further improve the efficiency.
%\subsubsection{Runtime Environment }\label{sec:emulation}
The CEDR ecosystem allows integrating compile-time application analysis with a Linux-based runtime system. We choose CEDR over other runtime frameworks ~\cite{augonnet2011starpu, donyanavard_sparta_2016, donyanavard_sosa_2019, maity_2021_SEAMSSelfOptimizing, martins_hierarchical_2019,tan_picos++_2019,  moazzemi_2019_HESSLEFREEHeterogeneous} as it enables compilation and development of user applications for heterogeneous SoCs, evaluating the performance of pre-silicon heterogeneous hardware configurations based on dynamically arriving workload scenarios through distinct plug-and-play integration points in a unified workflow.

% \subsection{Task definition}\label{sec:Registration}
% Our system processes tasks of two types:
% \begin{itemize}
%     \item \textbf{Type-1:} These tasks have low complexity and can be run on CPU in single thread. They do not require any change on the user side.
%     \item \textbf{Type-2:} These tasks are computation intensive. They can be run on existing accelerators or in parallel using multi-threading on CPU. Users are required to register Type-2 tasks with C-models and interface functions to invoke the driver of accelerators if they are for accelerators. Here the C-model is flexible enough that users can either use third-party libraries or just manually implement them. 
% \end{itemize}
%Memory analysis generates a load memory tuple set and a store memory tuple set for each node as illustrated in the bottom right figure with green and red bars. In the figure the y-axis shows id of each node, the x-axis shows the address range accessed by each node in a normalized form. Other than start and end addresses, each memory tuple contains memory footprint information, number of memory accesses and the task label of the node. For example the node 1 is an FFT, that is a 2048 point FFT with a memory footprint of 16384 bytes.

%\subsection{\sout{Running Example}}
\subsection{Program Model}
Our system partitions the user application into two types of tasks to represent the execution flow as  \emph{Type-1 Task} and \emph{Type-2 Task} regions as described below. 

\subsubsection{Type-1 Task} These tasks represent code segments that have low computation complexity, function as auxiliary, and are suitable for executing on the CPU core in a single thread. 

\subsubsection{Type-2 Task} These tasks represent compute-intensive segments of the execution at the task level such as FFT and matrix multiplication. We assume that there is accelerator support for \emph{Type-2 Tasks} and they can be executed in parallel using multi-threading on CPU or on the accelerator depending on the state of the system resources. Users are required to register Type-2 tasks with C-models and interface functions to invoke the driver of any supported accelerators. Here the C-model is flexible enough that users can either use third-party libraries or just manually implement them.
\subsection{Running Example}
We utilize a running example that consists of commonly used program structures with a representative computation task in our problem domain, namely, FFT(Fast Fourier Transform).  Figure~\ref{fig:code-example} presents the workings of our tool flow for the user application. Figure~\ref{fig:code-example}a is crafted in such a way that it includes features where static analysis is not feasible for detecting data dependence such as the use of pointers for passing variables (lines 2 and 8), function pointers for tasks (lines 6 and 11),  different function spaces while calling the tasks (lines 18 and 20), and loop iteration counter that is not known at compile time (line 19). Pre-processing step refactors the user application through inlining functions dynamically, redirecting function pointers and flattening loops that iterate over \emph{Type-2 Tasks}. This prepares the application for data DAG generation that captures concurrent execution paths over \emph{Type-2 Tasks} through dynamic memory analysis followed by control and data flow analysis. Finally refactored code is instrumented with pthread insertion for runtime system to be able to schedule the detected parallel Type-2 Tasks concurrently. 
In the following subsections, we present the details of our approach by walking through the steps of transforming a user application written in C/C++ into a representation that allows the runtime system to execute tasks without dependencies in parallel.

%\liang{
%The sample input code in Figure~\ref{fig:code-example}a has the listed scenarios where static analysis is not feasible for data dependence detection. 
%Dynamic profiling based preprocessing will help guide to flatten the loop and redirect the function pointers with identified tasks whose result is shown in Figure~\ref{fig:code-example}b. After that, the parallel code generation process will generate the parallelized code as shown in Figure~\ref{fig:code-example}c. }
\subsection{\textbf{Step 1}: Pre-processing}
\label{sec:preprocess}

%Typical software-defined radio applications written in C++ have code structures that make them hard to perform code analysis and transformation, as illustrated in Figure~\ref{fig:code-example}a. These structures include the use of pointers for variable passing, the use of function pointers to specify the tasks, the use of different function spaces while calling the tasks, and the unknown task loop iteration number. Note that recursion is not covered because we hardly see it in the radio applications yet.
In this step, we implement a profile-guided method to flatten the loops and inline the functions that interface with \emph{Type-2 Tasks}. The pre-processing step parses the user application and automatically instruments it with functions as illustrated in Figure~\ref{fig:code-example}b (lines 3, 15, 18 and 20) to be able to trace each Type-2 Task and loop identifier. The instrumented user application is compiled and executed to collect profiling data on loop trip count and function pointer destination. The pre-processing step then parses the instrumented user application second time, flattens the loop and redirects Type-2 Task calls. In our running example, assuming that loop trip count is two the output of the pre-processing step is illustrated in Figure~\ref{fig:code-example}c.

\subsection{\textbf{Step 2}:Data DAG Generation}
\label{sec:data-dag}

Implementation flow for this step is illustrated in Figure~\ref{fig:ToolFlow}b  where we start with compiling the preprocessed application using Clang to generate the LLVM IR. Recall that a user application is represented with a \emph{Control DAG}, where each node represents either \emph{Type-1} or \emph{Type-2} tasks and the edges between the nodes represent the flow of the execution. Here a task  can contain multiple basic blocks. We trace basic block control related flags such as \texttt{BasicBlockEnter}, \texttt{BasicBlockExit} along with task control related flags such as \texttt{TaskEnter} and \texttt{TaskExit} using TraceAtlas~\cite{uhrie2020automated} in \emph{Dynamic Profiling} stage. We generate the \emph{Control DAG} structure of the application based on the above four control flags. 
In our running example, the serial flow of program execution (\emph{ReadData(a0), FFT(a0,b0,.,.), WriteData(b0), ReadData(a1), FFT(a1,b1,.,.), WriteData(b1)}) illustrated in Figure~\ref{fig:code-example}c corresponds to the \emph{Control DAG} with six nodes  ($0\rightarrow1\rightarrow2 \rightarrow3\rightarrow4 \rightarrow5$) shown in Figure~\ref{fig:ToolFlow}b. The entry and exit points collected at the task level allow for segmenting the program into Type-1 and Type-2 tasks. Each time \texttt{TaskEnter} or \texttt{TaskExit} is seen, a new node is pushed into the Control DAG, and the basic blocks that appear between \texttt{TaskEnter} and \texttt{TaskExit} are recorded as the basic block elements of the node. The traced memory instructions in TraceAtlas include \texttt{LoadAddress}, \texttt{StoreAddress}, along with a set of LLVM intrinsic memory instructions such as \texttt{MemCpy}, \texttt{MemSet}, and \texttt{MemMove}. The addresses of memory instructions, control flags, and \emph{Control DAG} are used during \emph{Memory Analysis} stage to detect data dependencies and expose parallelism among the tasks.

The \emph{Memory Analysis} parses the \emph{Control DAG} and uses the trace information collected from \emph{TraceAtlas} to generate the load/store memory tuple sets for each node. We define a \emph{load memory tuple set} and a \emph{store memory tuple set} for each task of the Control DAG to represent the continuously accessed memory space for read and write activities, respectively. Each tuple for a task is composed of \emph{start address}, \emph{end address}, \emph{number of memory accesses}, \emph{number of bytes accessed}, and \emph{task label}. The memory tuples are stored in a red-black tree structure with the start address as the key to reducing the indexing time.

\begin{figure}[t]
\centerline{\includegraphics[width=0.98\columnwidth]{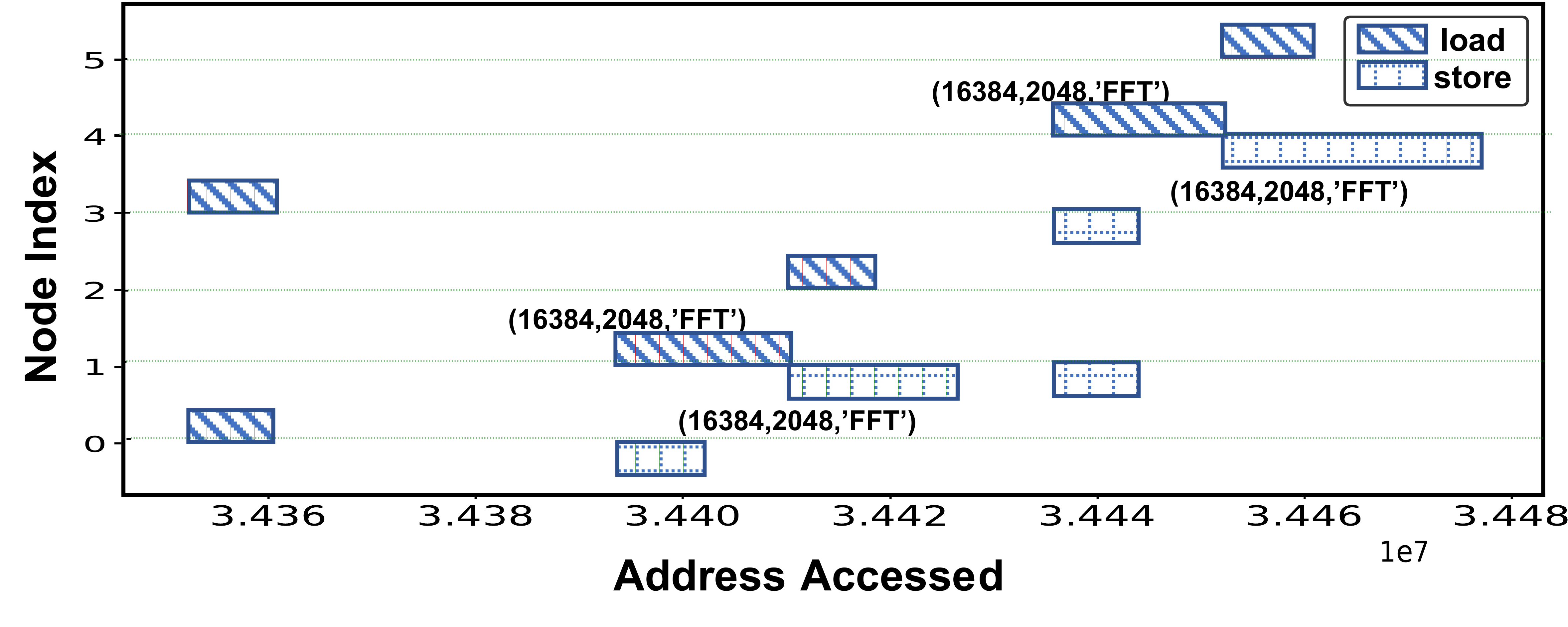}}
\vspace{-2mm}
\caption{ Memory Dependence Analysis. The address range that a node accesses when reading and writing are shown with the load and store tuple set. When the load tuple set of the successor does not overlap with the store tuple set of the predecessor, they do not have a data dependency.}
\vspace{-2mm}
\label{fig:memory-analysis}
\end{figure}

\begin{figure}[t]
\centerline{\includegraphics[width=1\columnwidth]{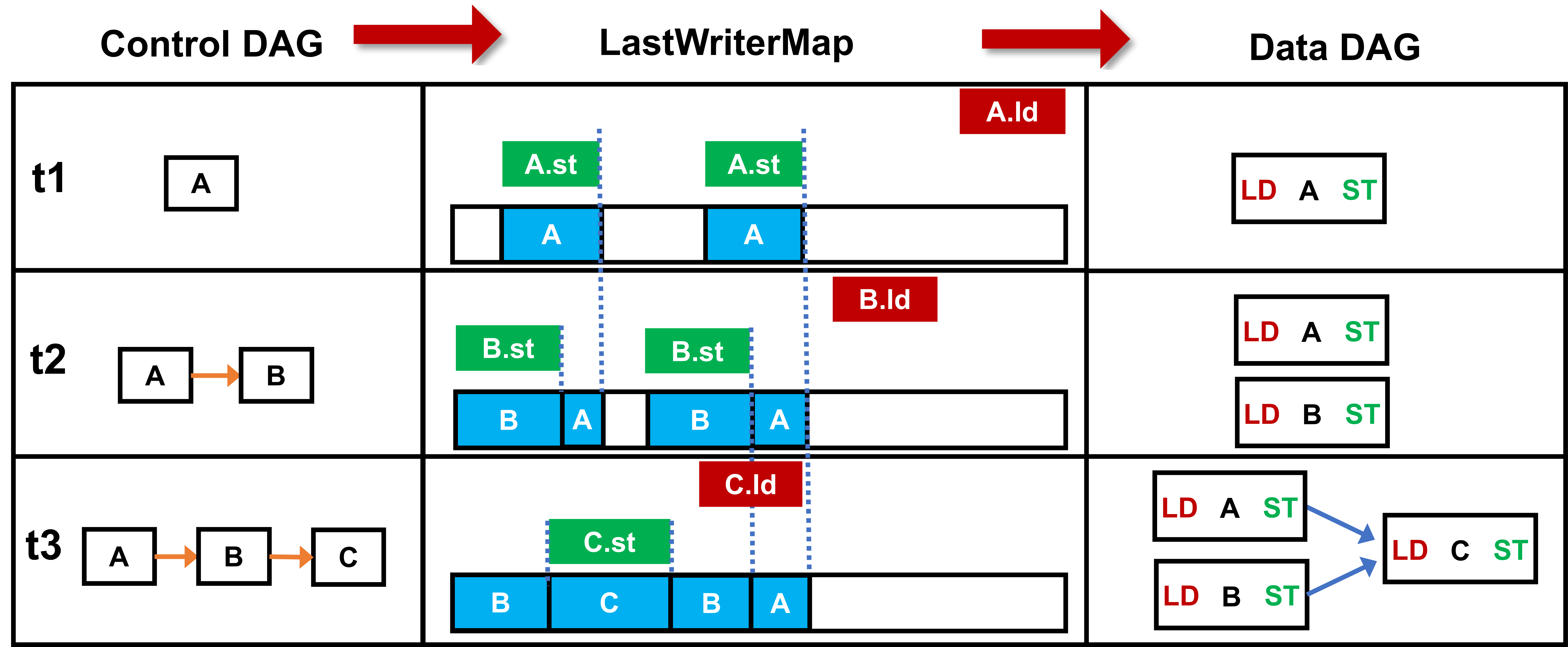}}
\caption{
Step by step (\textbf{t1} to \textbf{t3}) \emph{Data DAG} Generation for \emph{Control DAG} of $A \rightarrow B \rightarrow C$ scenario.
% Initially \texttt{LastWriterMap} is empty. Store tuple set of A is directly written into the \texttt{LastWriterMap} and points to A. Load tuple set of A has no overlap with existing tuples in the \texttt{LastWriterMap}. No edge is added. At \textbf{t2}, store tuple set of B overwrites part of the existing memory tuples that point to A and change the pointed value to B. Also no edge is created because B's load tuple set has no overlap with the \texttt{LastWriterMap}. At \textbf{t3}, C modifies the \texttt{LastWriterMap}. Load tuple set of C overlaps with memory tuples in the \texttt{LastWriterMap} that point to both A and B. Edges are created from A to C and B to C completing the data DAG generation.
\vspace{-4mm}
}
%\sout{\emph{Data DAG} Generation for a scenario where Task C depends on two independent Tasks A and B: \textbf{t1}, \textbf{t2}, and \textbf{t3} are timestamps. At \textbf{Timestamp t1}, the \texttt{LastWriterMap} is empty. Node A is parsed. The store tuple set of A is directly written into the \texttt{LastWriterMap} and points to A. The load tuple set of A has no overlap with existing tuples in the \texttt{LastWriterMap}. No edge is added. At \textbf{\sout{Timestamp} t2}, B comes and its store tuple set overwrites part of the existing memory tuples that point to A and change the pointed value to B. Also no edge is added because B's load tuple set still has no overlap with the \texttt{LastWriterMap}. At \textbf{Timestamp t3}, C comes and changes the \texttt{LastWriterMap} in the same way. By checking the load tuple set of C, we see overlaps with memory tuples in the \texttt{LastWriterMap} that point to both A and B. Then edges are added from A to C and B to C. All nodes in the control DAD are processed, the data DAG generation is done}

\label{fig:DAGGen}
\end{figure}

The visual representation of memory analysis is shown in Figure~\ref{fig:memory-analysis}, where the y-axis shows an index of each node in the \emph{Control DAG}, x-axis shows the address range accessed (load or store) by each node through labeled rectangles in a normalized form. 
For example, node index one is the first FFT task in the user application with 2048 samples and 16,384 bytes of memory footprint starting with the load address range followed by the store address range corresponding to read and write operations respectively. The read after write dependence is identified by checking if the load tuple set of the successor node overlaps with the store tuple set of the predecessor node. A data dependency between predecessor and successor nodes exists only if the predecessor node is the last node to write into the load address space of the successor node. In the running example, referring to the memory analysis result in Figure~\ref{fig:memory-analysis}, 
Nodes 1 and 3 (\emph{FFT()} and \emph{ReadData(a1)}) have the same store address space. 
Based on the order of execution, the true dependency is between Nodes 3 and 4 (second FFT task). 
The second FFT task (Node 4) has a load address space that does not overlap with the store address space of the first FFT task (Node 1). 
Finally, we implement the \texttt{LastWriterMap} data structure as illustrated in Figure~\ref{fig:DAGGen} to keep track of the address spaces modified by each node and utilize this representation to identify Type 2 Tasks that can be executed in parallel.

The \emph{Data DAG} stage in this phase detects \emph{Type-2 Tasks} that can be executed in parallel and generates the corresponding \emph{Data DAG}. The \emph{Data DAG} is built on the \emph{Control DAG} structure, with each edge representing the data dependencies between the tasks.

%The output of the \emph{Memory Analysis} along with the \emph{Control DAG} are then processed to determine whether Type-2 tasks can be executed in parallel or not, and generate the \emph{Data DAG}.  
%After all the nodes in the \emph{Control DAG} are parsed, the final \emph{Data DAG} is generated.

%A task can load from an address range that has been written by several predecessor tasks, but the successor task will only depend on the task that is the last one to write to the address range. 
%\sout{We parse the \emph{Control DAG} from first node to last node. We implement the \texttt{LastWriterMap} data structure as illustrated in Figure~\ref{fig:DAGGen} to keep track of the node that is modifying a given address range.} 

%, whose key is the pointer to a memory tuple and value is the node index of the task that is the last one to write into that memory tuple. 
%This data structure is maintained to track the address space modified by each node. 
Figure~\ref{fig:DAGGen} shows state of the \texttt{LastWriterMap} and \emph{Data DAG} generation in three steps ($t1$ to $t3$) for a simple scenario composed of independent Tasks A and B feeding their data to Task C with a \emph{Control DAG} of $A \rightarrow B \rightarrow C$.
%\sout{how the nodes in the \emph{Control DAG} are parsed one by one in three steps ($t1$ to $t3$), and how the \texttt{LastWriterMap} is maintained during each step, and finally used to generate the \emph{Data DAG}.} 
Each time a new node is visited in the \emph{Control DAG}, its store memory tuple set is written into the \texttt{LastWriterMap}. 
We overwrite the existing memory tuples and change the pointed value to the new node's index if the store memory tuple set of the new node overlaps with the store memory tuple set maintained in the current \texttt{LastWriterMap}. 
At the end of the second step, the {Data DAG} with two independent nodes are generated as load address space of \emph{Task B} ($B.ld$) is not overlapping with the store address space in the \texttt{LastWriterMap}. 
On the other hand, in step 3, edges from \emph{Task A} and \emph{Task B} to \emph{Task C} are generated due to the load-store overlap between $C.ld$ and \emph{store memory tuple} in the \texttt{LastWriterMap} that points to both \emph{Task A} and \emph{Task B}. 

\begin{figure}[t]
\centerline{\includegraphics[width=0.8\columnwidth]{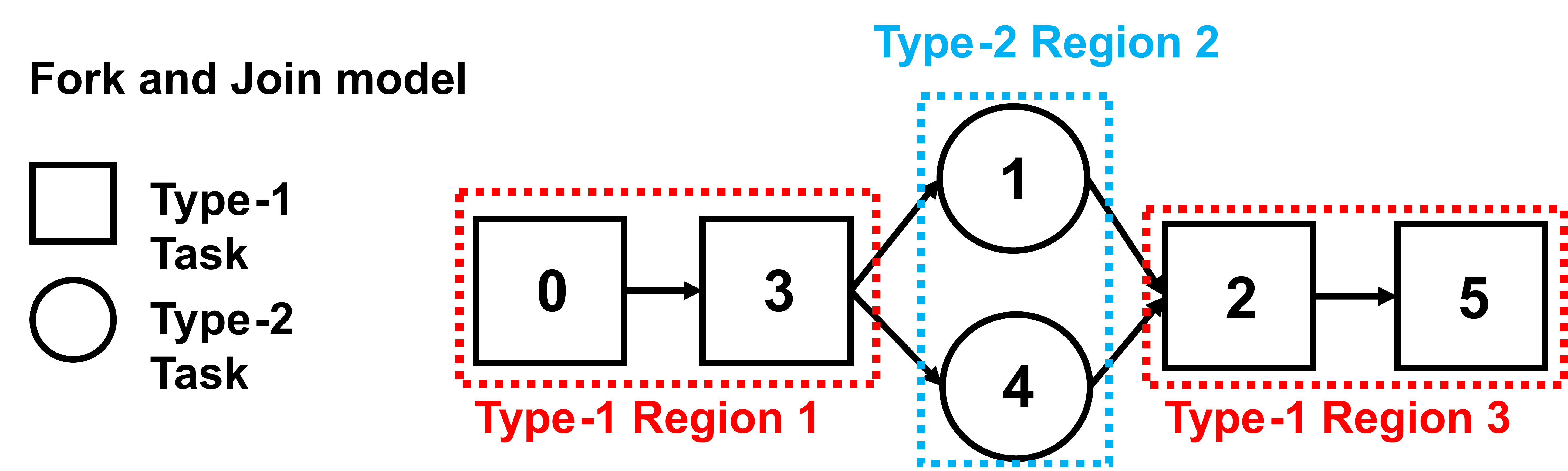}}
\caption{
%\sout{Task Reordering: The application has two independent Type-2 FFT tasks with distinct IOs. In DAG representations, Node 1 and 4 represent Type-2 tasks that are FFTs, Node 0 and 3 represent Type-1 tasks that read input data from a file, Node 2 and 5 represent Type-1 tasks that write output data to a file. From the data DAG, we can see there are two independent paths. Since we only want to parallelize Type-2 tasks, the execution should follow the fork and join model in step 3. The algorithm in Fig.\ref{fig:schedule} is used to generate a schedule according to which the control DAG can be correctly reordered. The program is then transformed to follow the reordered control DAG shown in step 4.   }
Task Reordering for the running example from Figure~\ref{fig:code-example}: Nodes 1 and 4 represent two independent FFTs (Type-2 Tasks) with distinct IOs, Nodes 0 and 3 represent reading input data (Type-1 Tasks), and Nodes 2 and 5 represent writing output data (Type-1 Tasks). The algorithm in Figure~\ref{fig:schedule} generates a schedule with Control DAG nodes reordered following the fork and join model. }
\vspace{-6mm}
\label{fig:taskReord}
\end{figure}

%After identifying the tasks that can be executed in parallel, we transform the user application to be represented as series of code sections composed of Type-1 and Type-2 regions successively where a Type-2 region absorbs all independent Type-2 tasks. We implement the \emph{Schedule Generation} algorithm for this transformation and generate a \emph{Schedule DAG} that follows the fork and join parallel model as illustrated in Figure~\ref{fig:taskReord} for our running example shown in Figure~\ref{fig:ToolFlow}. This representation is our interface to the runtime system for the parallel tasks in each Type-2 region to be scheduled concurrently. The \emph{Schedule Generation} algorithm is realized through seven steps as illustrated in Figure~\ref{fig:schedule} using \emph{Data DAG} and \emph{Control DAG} as its inputs. We define a task as a ready task if all of its parent tasks in the \emph{Data DAG} have been scheduled. Steps 1-3 generate \emph{Schedule DAG} for Type-1 Tasks (Nodes 0 and 3) and mark these nodes as visited. Steps 4-6 updates \emph{Schedule DAG} with Type-2 Tasks (Nodes 1 and 4) and mark these nodes as visited. All steps are repeated by visiting Type-1 Tasks followed by visiting Type-2 Tasks that remain in the DAG until all DAG nodes have been visited.

\begin{figure}[t]
     \centering
     \begin{subfigure}[b]{0.47\textwidth}
         \centering
         \includegraphics[width=1\columnwidth]{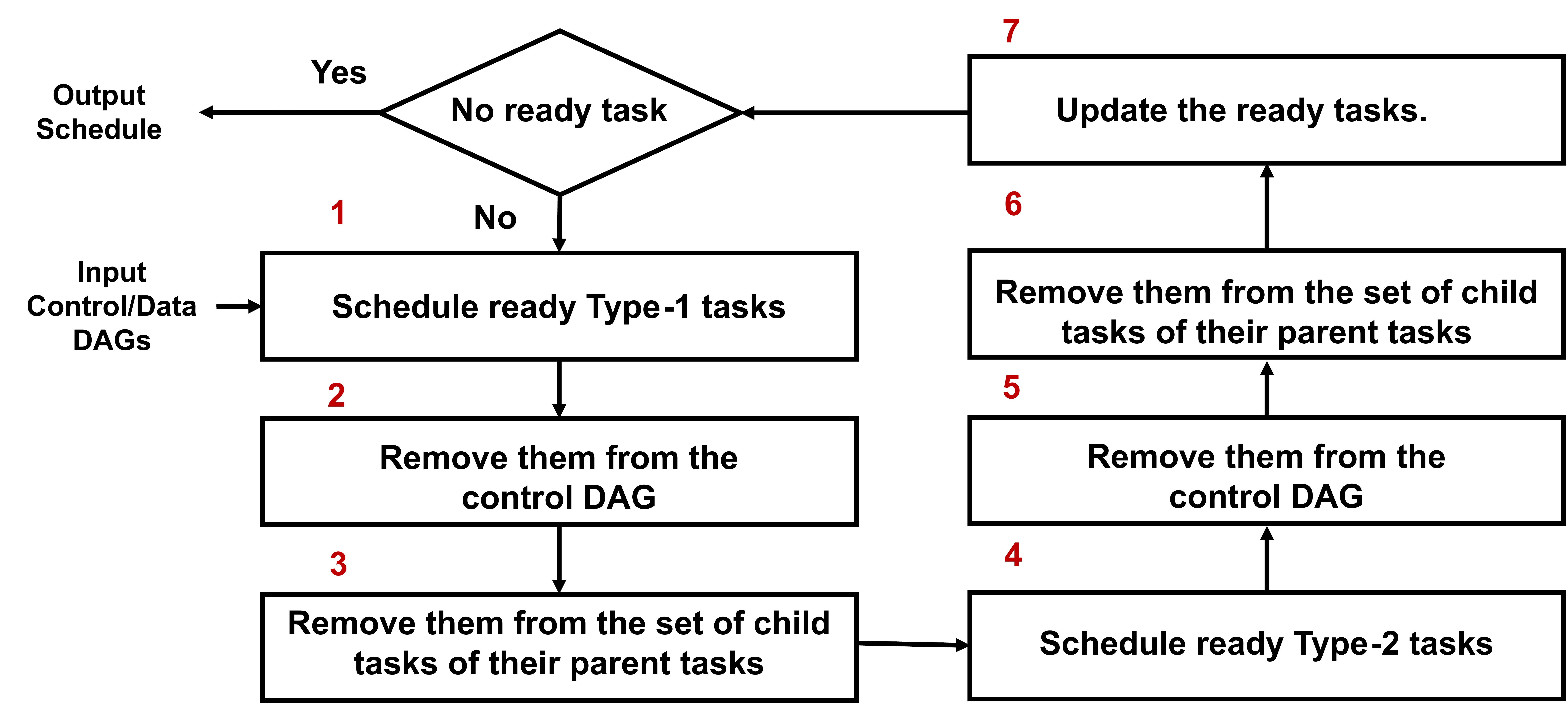}
         \caption{Schedule Generation algorithm flow.}
         \label{fig:schedAlg}
     \end{subfigure}
     \hfill
     \begin{subfigure}[b]{0.4\textwidth}
         \centering
         \includegraphics[width=1\columnwidth]{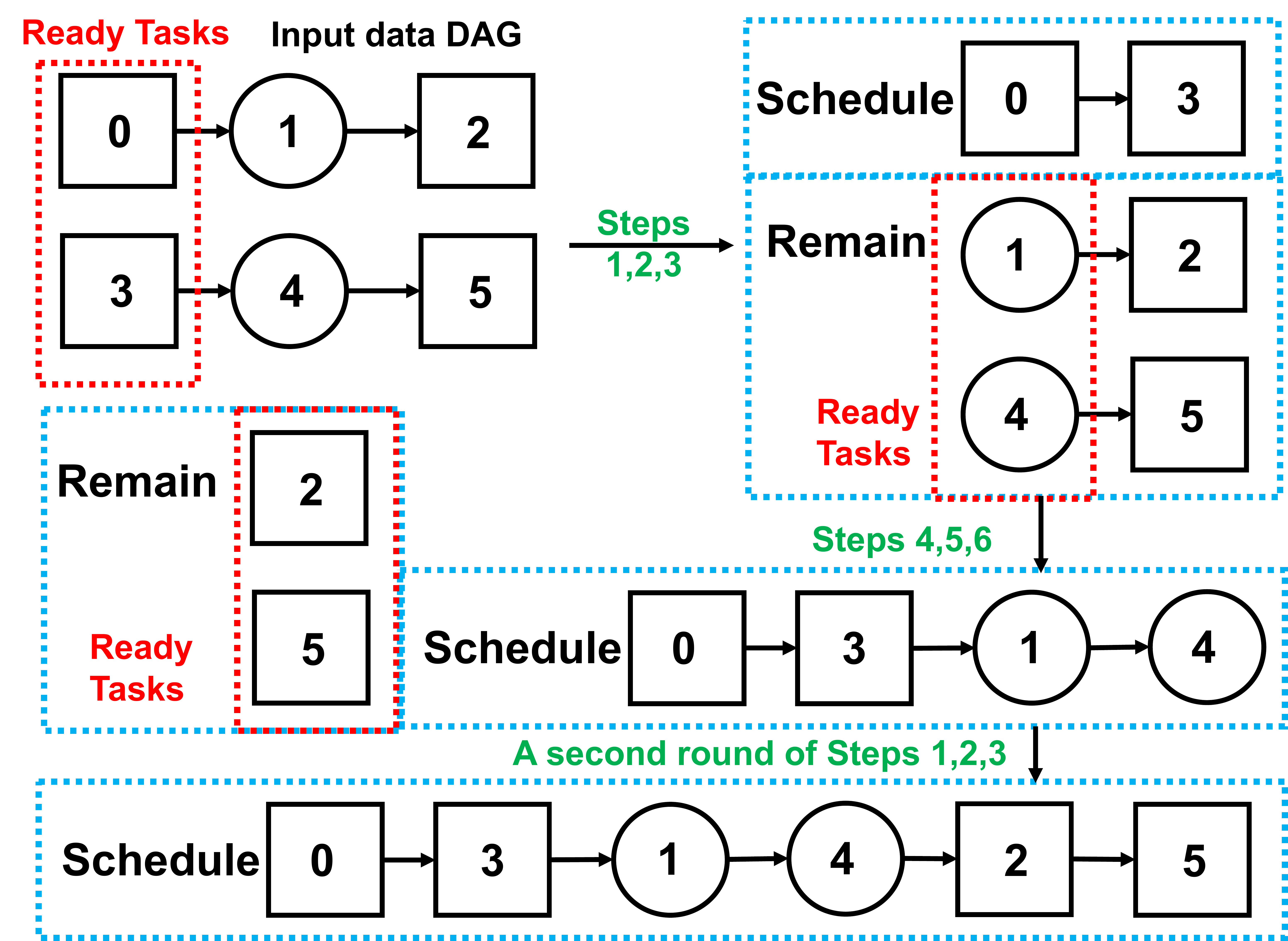}
         \caption{State of the \emph{Schedule} as Schedule Generation progresses.}
         \label{fig:schedExm}
     \end{subfigure}
     \caption{
         The Schedule Generation algorithm reorders DAG nodes for the running example from Figure~\ref{fig:code-example}. 
        }
\vspace{-2mm}
     \label{fig:schedule}
\end{figure}

\subsection{\textbf{Step3:} Parallel Code Generation}
\label{sec:codegen}

We define the group of Type-1 tasks that can be scheduled for execution before a Type-2 task as the \emph{Type-1 Region} and the group of Type-2 tasks that can be scheduled before a Type-1 task as the \emph{Type-2 Region}. After identifying the Type-2 tasks that can be executed in parallel, we transform the user application to be represented as a series of code sections composed of Type-1 and Type-2 regions successively in \emph{Task Reordering} stage. We implement the \emph{Schedule Generation} algorithm for this transformation and generate a \emph{Schedule DAG} that follows the fork-and-join parallel model as illustrated in Figure~\ref{fig:taskReord} for our running example shown in Figure~\ref{fig:code-example}. The \emph{Schedule Generation} algorithm is realized through seven steps as illustrated in Figure~\ref{fig:schedule} using \emph{Data DAG} and \emph{Control DAG} as its inputs. We define a task as a ready task if all of its parent tasks in the \emph{Data DAG} have been scheduled. Steps 1-3 generate \emph{Schedule DAG} for Type-1 Tasks (Nodes 0 and 3) and mark these nodes as visited. Steps 4-6 update \emph{Schedule DAG} with Type-2 Tasks (Nodes 1 and 4) and mark these nodes as visited. All steps are repeated by visiting Type-1 Tasks followed by visiting Type-2 Tasks that remain in the DAG until all DAG nodes have been visited. 
While our running example in Figure~\ref{fig:code-example}c involves two parallel Type-2 Tasks, the \emph{Schedule Generation} algorithm implements a generalized solution that is capable of clustering \emph{N} parallel Type-2 Tasks into a single region to realize \emph{N-way} parallelism.

%\josh{
%Include a statement about generalizability of the approach. ("This is just segmentation, by recognizing that there are N distinct overlap regions, we can see that this approach is trivially generalizable to N-way parallelization through adjustments to the counter threshold")
%}

% \begin{figure}[t]
% \centerline{\includegraphics[width=0.8\columnwidth]{images/cedr.png}}
% \caption{CEDR management flow }
% \vspace{-4mm}
% \label{fig:cedr}
% \end{figure}

\begin{figure}[t]
     \centering
    %  \begin{subfigure}[b]{0.15\textwidth}
    %      \centering
    %      \includegraphics[width=1.25\columnwidth]{images/pthread.png}
    %      \caption{Serial execution of two kernels.}
    %      \label{fig:pthread}
    %  \end{subfigure}
    %  \hfill
     \begin{subfigure}[b]{0.23\textwidth}
         \centering
         \includegraphics[width=0.84\columnwidth]{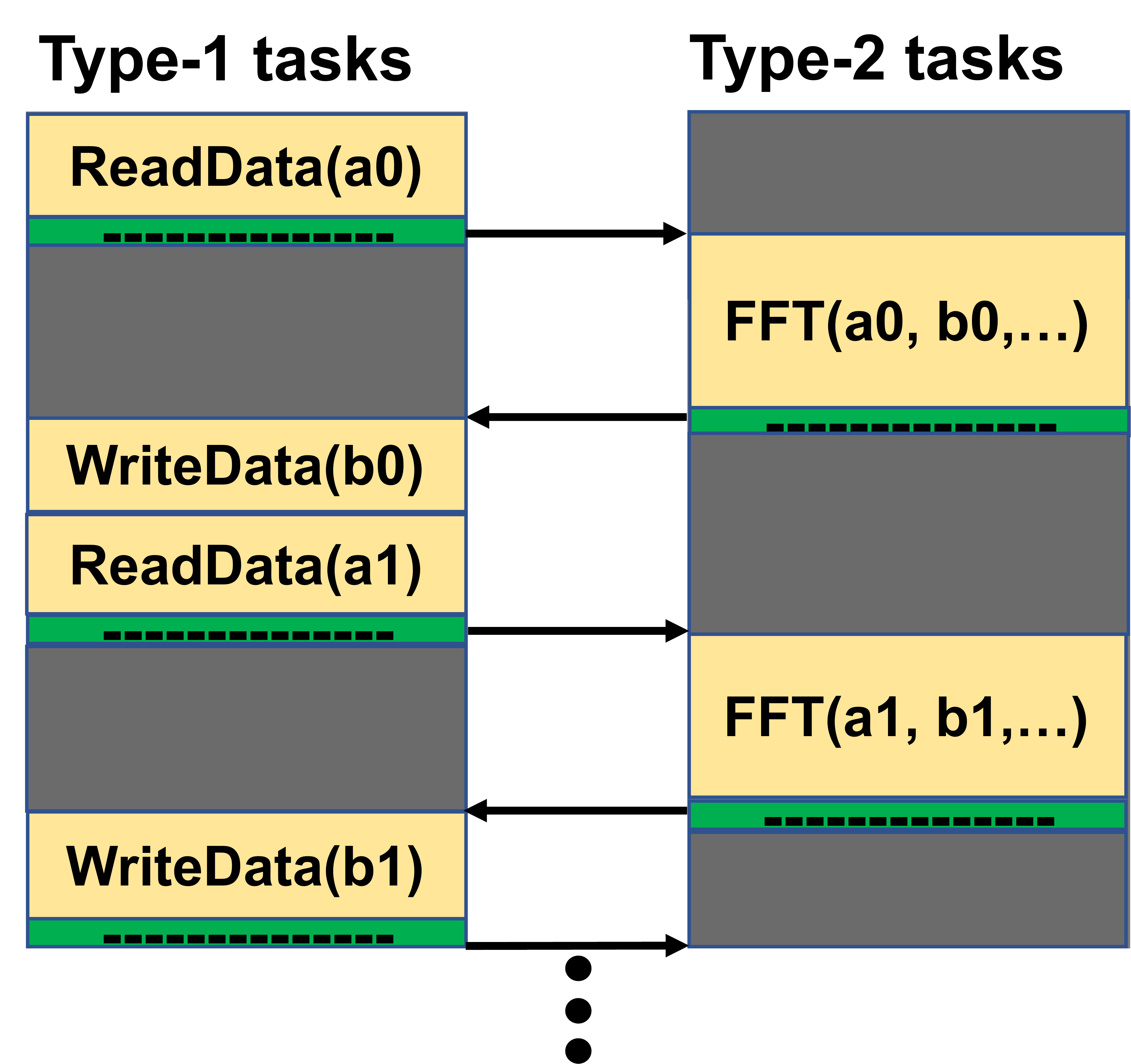}
         \caption{Type-2 tasks execute serially.}
         \label{fig:serial}
     \end{subfigure}
     \begin{subfigure}[b]{0.23\textwidth}
         \centering
         \includegraphics[width=0.9\columnwidth]{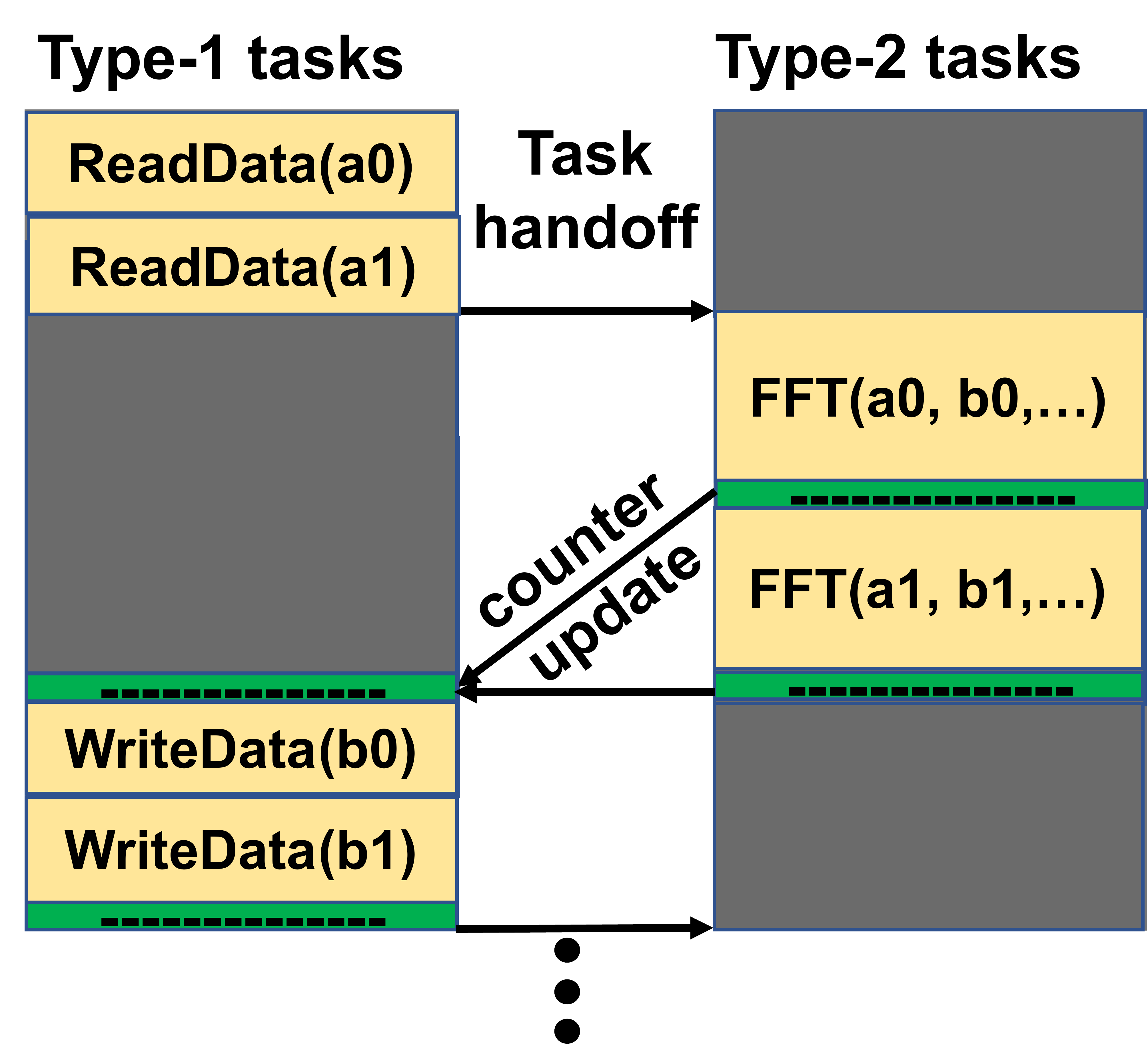}
         \caption{Tasks rearranged and merged.}
         \label{fig:counter}
     \end{subfigure}
     \hfill
     \caption{
         Code transformation for parallel execution through task merging. 
        }
\vspace{-4mm}
     \label{fig:kernelrep}
\end{figure}

Referring to the user application shown in Figure~\ref{fig:code-example}c, during the application refactoring process of \emph{Pthread Insertion}, all Type-2 Tasks (FFTs) are swapped with the \textit{enqueue kernel} as shown in the tool-generated code with Figure~\ref{fig:code-example}d. This representation is our interface for the Type-2 tasks to be scheduled by the runtime resource manager. CEDR operates as a background \emph{Daemon Process} in the Linux user space and applications are submitted through inter-process communication (IPC). The CEDR \emph{Management Thread} parses the dynamically arriving application binaries and through the \textit{enqueue kernel} function places Type-2 Tasks, whose dependencies are resolved, into the ready queue. The scheduler makes task-to-PE mapping decisions for those tasks that are in the ready queue. Then CEDR launches a worker thread for each Type-2 Task to execute on the selected PE and monitors the state of execution through \emph{PE Tracker}.    

The trace-based analysis and \emph{Data DAG} representation enable specifying that the user application in Figure~\ref{fig:code-example} can be refactored from the original structure shown in Figure~\ref{fig:serial} into Type-1 and Type-2 regions as illustrated in Figure~\ref{fig:counter}. 
Along with this transformation, the run time environment needs to know the number of independent Type 2 Tasks grouped together into a single region. For this, during \emph{Pthread Insertion} stage, code refactoring also involves automatically inserting \texttt{pthread} initialization and \texttt{conditional wait statements} into the transformed application. In this final code representation, two enqueue kernels are placed one after another without barrier synchronization but there is a while loop waiting for those two FFT functions to be completed. At run time, the OS thread initializes the counter value to 0 and waits for the counter to meet the condition. For our running example, the conditional wait value is set to two.
We updated CEDR to support the execution of parallel \emph{Type-2 Tasks} in \emph{Type-2 Region}.
Given that the data initializations are completed by the OS thread, CEDR places both FFTs into the ready queue as there is no barrier between the two FFTs. 
The scheduler then picks up both FFTs from the ready queue and makes the task to PE mapping decisions. For each FFT, a worker thread is launched and both functions get executed in parallel.  Through the \emph{PE Tracker}, the CEDR management thread monitors the execution of each FFT. As soon as an FFT task gets completed, the CEDR worker thread increments the counter value by one and passes this information back to the OS thread. When the counter reaches to two, the OS thread resumes its Type-1 region possibly to the next phase that needs outputs of the two FFTs.
The runtime system supports executing \emph{N-way} parallelism as the counter value is a parameter that is generated during the \emph{Pthread Insertion} stage.

\begin{figure}[b]
     \centering
     \includegraphics[width=0.9\columnwidth]{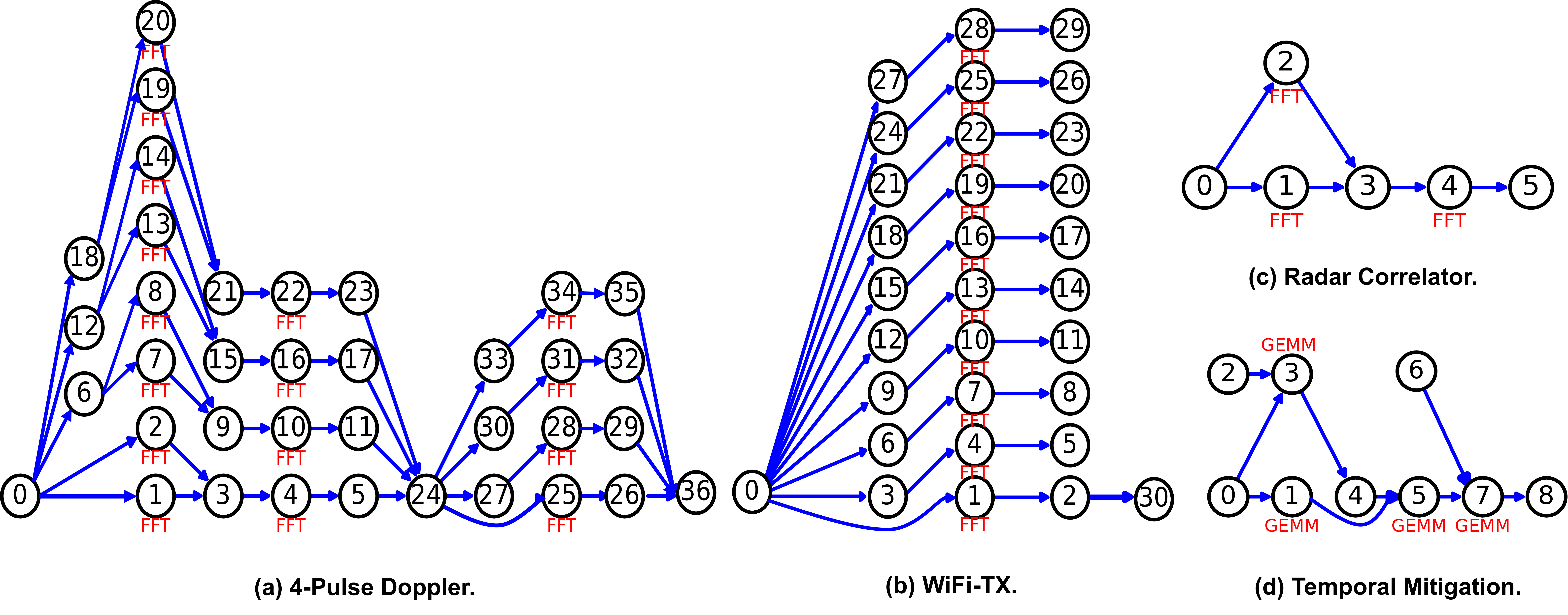}
     \caption{The tool generated \emph{Data DAG} of the test applications.}
     \label{fig:DAG-figure}
\end{figure}
     
\section{Experimental Setup}
\label{sec:setup}

\begin{table}[]
\centering
\begin{tabular}{|l|c|c|c|}
\hline
Application         & Type-2 task & Phases & Parallelism/Phase \\ \hline
4-Pulse Doppler     & FFT(256pt)         & 3           & 8, 4, 4                 \\ \hline
256-Pulse Doppler   & FFT(256pt)         & 3           & 512, 256, 256                \\ \hline
WiFi-TX             & FFT(128pt)         & 1           & 10                 \\ \hline
Radar Correlator    & FFT(512pt)         & 2           & 2, 1                  \\ \hline
Temporal Mitigation & GeMM(4x64)        & 2           & 2, 1                  \\ \hline
\end{tabular}
\caption{Benchmark applications with \emph{Type-2 task}, the number of \emph{Type-2 task} phases, and the maximum number of parallel \emph{Type-2 tasks} for each phase.}
\label{tab:benchmark}
\vspace{-6mm}
\end{table}

\subsection{Applications}
For our analysis, we use Pulse Doppler, WiFi-TX, Radar Correlator and Temporal Mitigation as real-world applications from the domain of software defined radio. These applications have varying levels of parallelization and help illustrate 
%2-way to 512-way parallel execution paths illustrating 
the generalizability of our approach. \\
{\it Pulse Doppler} determines both the distance of an object and its velocity based on a series of short radar pulses emitted, and the user application observes the shift in the frequencies of the return pulses with respect to the input pulse. \\
{\it WiFi TX} implements a WiFi transmit chain, generating a single packet with 64 bits of input data and scrambling, encoding, modulating,  followed by forward error correction. \\
{\it Radar Correlator} models the use of a radar pulse to determine distance to an object by looking at the time delay in the received pulse compared to the input pulse.\\
{\it Temporal Interference Mitigation} receives a signal consisting of low-energy radar signals combined with high-energy communications data and applies a technique known as \textit{successive interference cancellation} to cancel out the communications data and extract the radar signals for further processing. 

Figure~\ref{fig:DAG-figure} shows the execution phases as a data flow graph (DFG) for each application generated by our tool chain, where a phase is defined as a distinct \emph{Type-2 Region}.  For the sake of simplicity, we show DFG for the 4-Pulse Doppler version which has three execution phases, but in our evaluations, we also use the full scale implementation with 256-Pulses.
In Table~\ref{tab:benchmark} we summarize the type of task, the number of execution phases, and the degree of task-level parallelism observed during each execution phase.  
\subsection{Evaluation Platforms}
For our evaluations, we utilize three diverse platforms, namely, DS3~\cite{arda2020ds3}, an event-driven simulator,
%DS3~\cite{arda2020ds3} event-driven simulator along with two off-the-shelf platforms. 
a homogeneous architecture based on Intel(R) Xeon(R) CPU E5-2650 v3 multicore processor, and
Xilinx Zynq Ultrascale+ ZCU102 MPSoC development board.

DS3 simulates the execution of an application represented as a DAG over a user specified heterogeneous SoC configuration where the task to processing element mapping decisions are handled through its built-in Earliest Finish Time (EFT) scheduler. This environment serves as a suitable platform to estimate performance gains of the task-level parallelism extracted through our tool chain. 

On the Xeon(R) CPU, each application is processed through our compiler tool chain and then executed through the CEDR environment over 8-cores. This setup allows us to validate the end-to-end integrated compiler and runtime flow over a homogeneous architecture and evaluate the performance gain with respect to serial execution.

The Xilinx Zynq Ultrascale+ ZCU102 MPSoC development board is used to  emulate a heterogeneous SoC with 3 ARM CPU cores and 1 FFT accelerator. This setup serves three purposes. First, it allows validating our ability to compile a user application for execution on a heterogeneous SoC. Second, we demonstrate our ability to execute parallel FFT tasks in a single application across a pool of heterogeneous resources concurrently. 
Third, we demonstrate our ability to manage workload scenarios where multiple independent user applications arrive dynamically and parallel execution at both application and task levels are realized through our integrated compile and runtime flow. To facilitate data transfer to and from accelerators, we use direct memory access (DMA) blocks to move data between the host ARM core and the FFT accelerator via the AXI4-Stream protocol.
On the host side, we utilize \textit{udmabuf} to enable contiguous userspace-accessible buffers for transferring data to and from the hardware accelerators. A user application communicates with the accelerators by writing the data into a udmabuf buffer and a DMA engine is then configured to move data from this buffer into an accelerator for processing. This setup, while allowing experimentation on a heterogeneous hardware configuration, does not offer a realistic SoC representation since we are not emulating a dedicated NoC. Therefore we strictly use it for functional verification purpose rather than conducting realistic performance evaluations. 

For the FPGA evaluation we use the Radar Correlator and WiFi-TX. We process them through our compiler tool chain and submit them as single or 100 instance jobs where the CEDR management thread parses application binary, monitors system resources, schedules tasks based on Minimum Execution Time (MET), Round Robin (RR) and Earliest Finish Time (EFT) schedulers. We measure application execution time as the difference between the end of the last task and the start of the first task of an application, including the overhead of all scheduling decisions and data transfers to and from the accelerator in between.   

\section{Results and Analysis}
\label{sec:results}

\begin{figure*}[t]
     %\centering
     \begin{subfigure}[b]{0.33\textwidth}
         \centering
         \includegraphics[width=0.9\columnwidth]{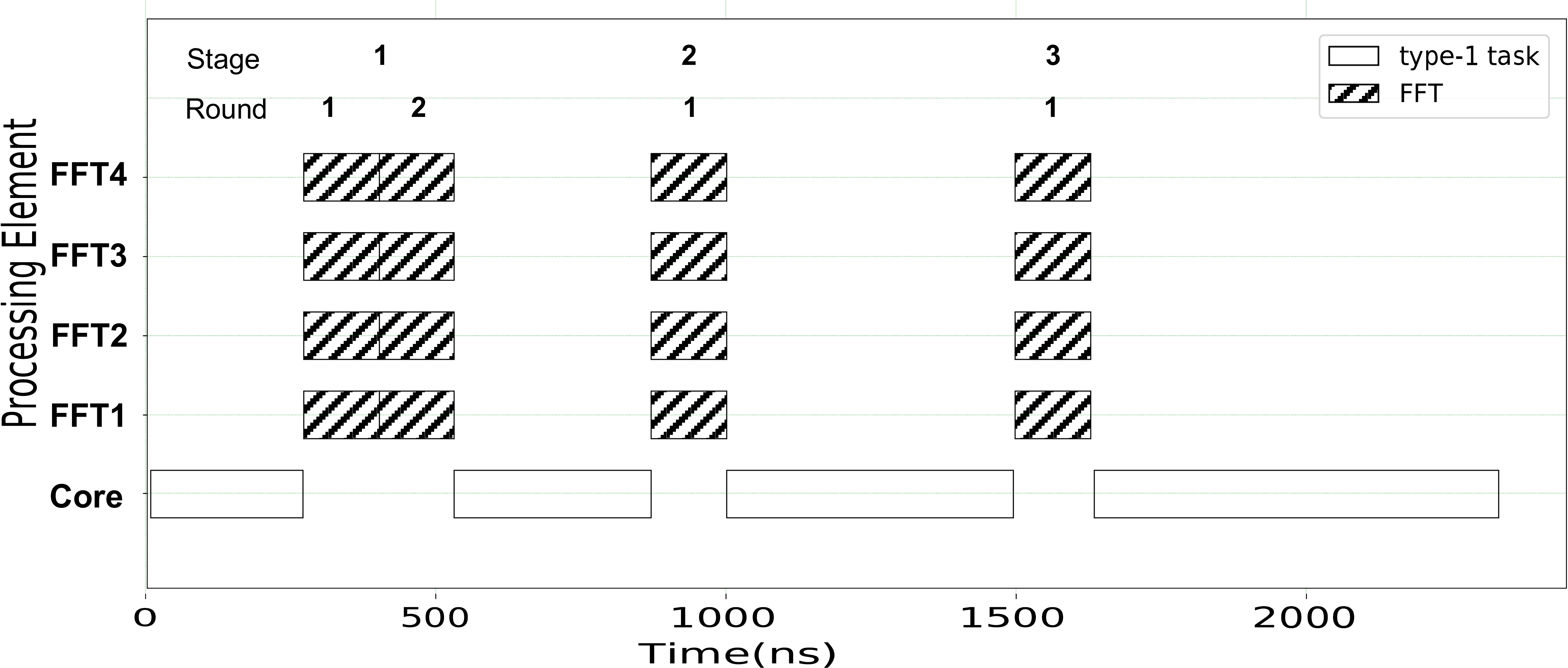}
         \caption{4-Pulse Doppler on SoC with 4 FFTs. }
         \label{fig:4-pulse-4fft}
     \end{subfigure}
     \begin{subfigure}[b]{0.33\textwidth}
         \centering
         \includegraphics[width=0.9\columnwidth]{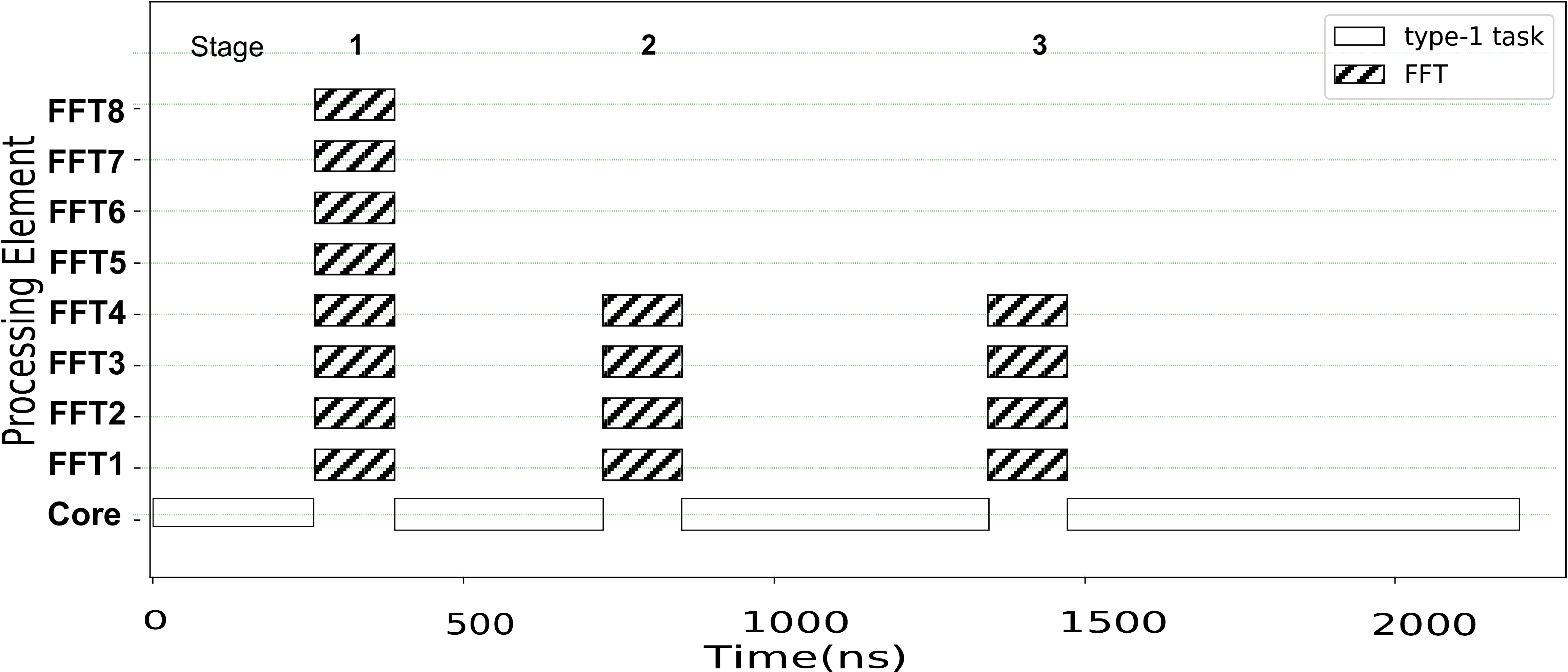}
         \caption{4-Pulse Doppler on SoC with 8 FFTs.}
         \label{fig:4-pulse-8fft}
     \end{subfigure}
    %  \vspace{-4mm}
    % \hfill     
     \begin{subfigure}[b]{0.33\textwidth}
         \centering
         \includegraphics[width=0.95\columnwidth]{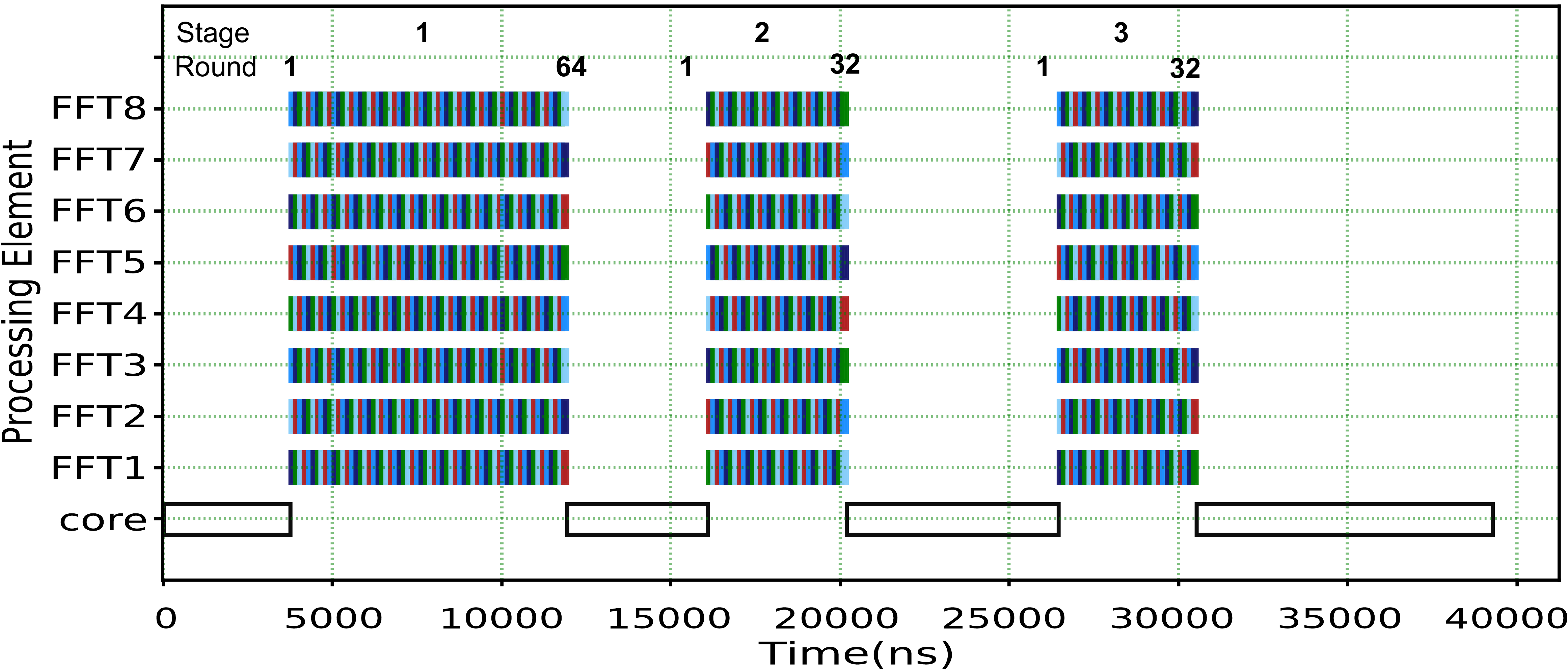}
         \caption{256-Pulse Doppler on SoC with 8 FFTs.}
         \label{fig:256-pulse-ds3}
     \end{subfigure}
     \caption{
          Execution flow for Pulse Doppler on DS3-simulated SoCs with varying numbers of FFT accelerators: (a) 8 parallel FFTs in the first stage are executed over 4 FFT accelerators in two rounds, followed by single round execution in second and third stages. (b) 8 parallel FFTs in the first stage are executed over 8 FFT accelerators in one round, followed by single round execution in the following two stages. (c) Full scale Pulse Doppler on an SoC with 8 FFTs where 256 FFTs in the first stage are evenly distributed as 64 FFT rounds over 8 cores followed by 32 FFT rounds in the subsequent two stages. 
        }
     \label{fig:pulse-gantt}
\vspace{-4mm}
\end{figure*}
\begin{figure}[t]
         \centering
         \includegraphics[width=0.9\columnwidth]{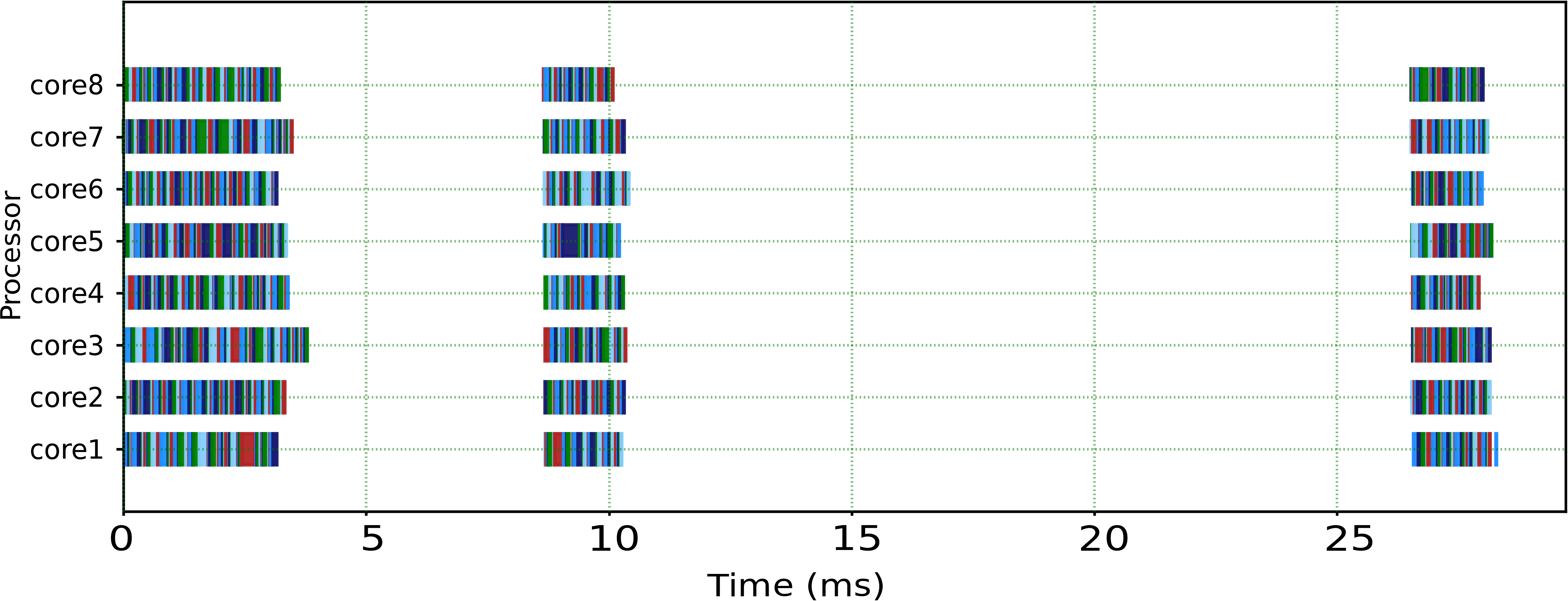}
     \caption{
          The execution flow of 256-Pulse Doppler in three stages over the 8-core Xeon processor managed by CEDR with 8 FFT accelerators matches the execution obtained from DS3 illustrated in Figure~\ref{fig:256-pulse-ds3}.% They show the same computation stage patterns. For the result on the processor, the first round of FFT execution on each core is dropped because, in order to load data from DRAM, it takes an extra long time compared with other rounds. It also counts all the environmental influences like the overhead of scheduling, OS management, cache locality, and so on. The SoC results don't count the environmental overheads. The execution time of type-1 tasks in SoC results is scaled since it is too large compared with the execution time of accelerators. Otherwise, it is hard to see the bar charts of the accelerator tasks. 
        }
     \label{fig:256-pulse-cpu}
     \vspace{-2mm}
\end{figure}

\begin{figure*}[t]
     \centering
     \begin{subfigure}[b]{0.33\textwidth}
         \centering
         \includegraphics[width=0.95\columnwidth]{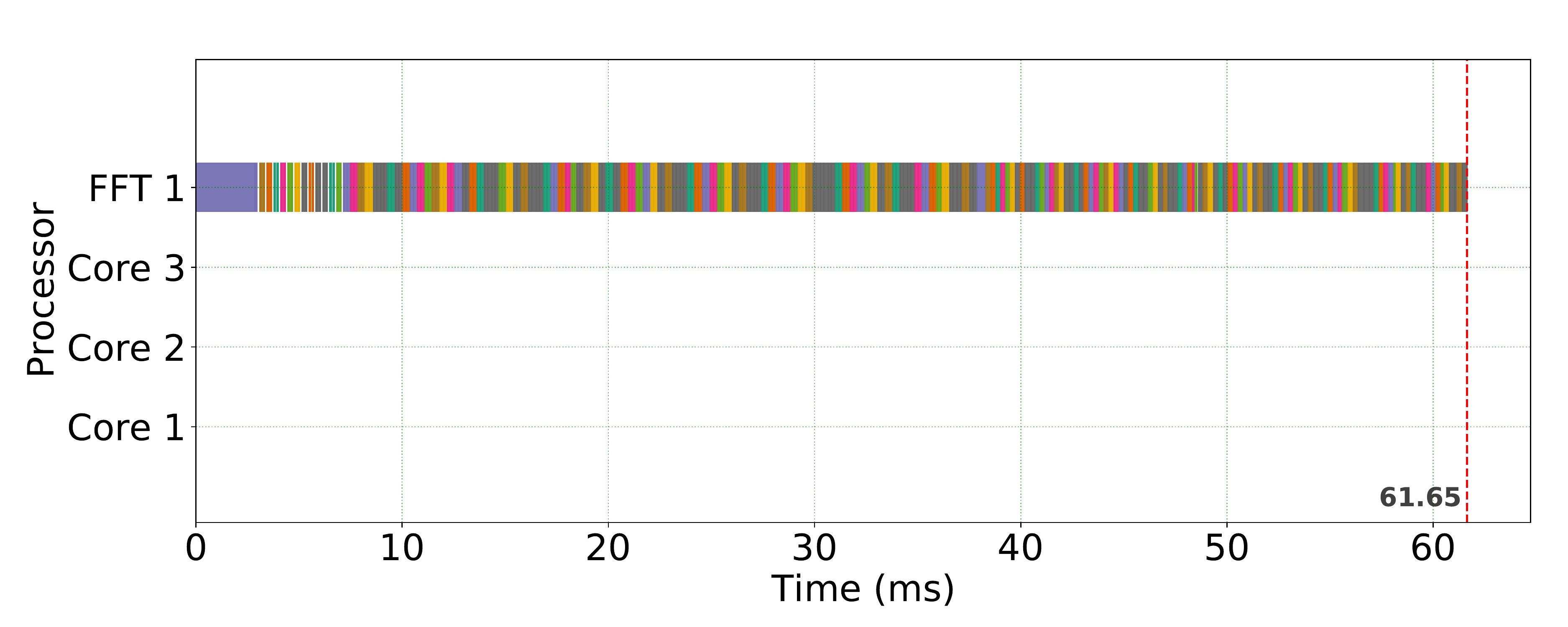}
         \vspace{-3mm}
         \caption{Minimum Execution Time - MET}
         \label{fig:parallel_radar_correlators_met}
     \end{subfigure}
     % \hfill
    \begin{subfigure}[b]{0.33\textwidth}
         \centering
         \includegraphics[width=0.95\columnwidth]{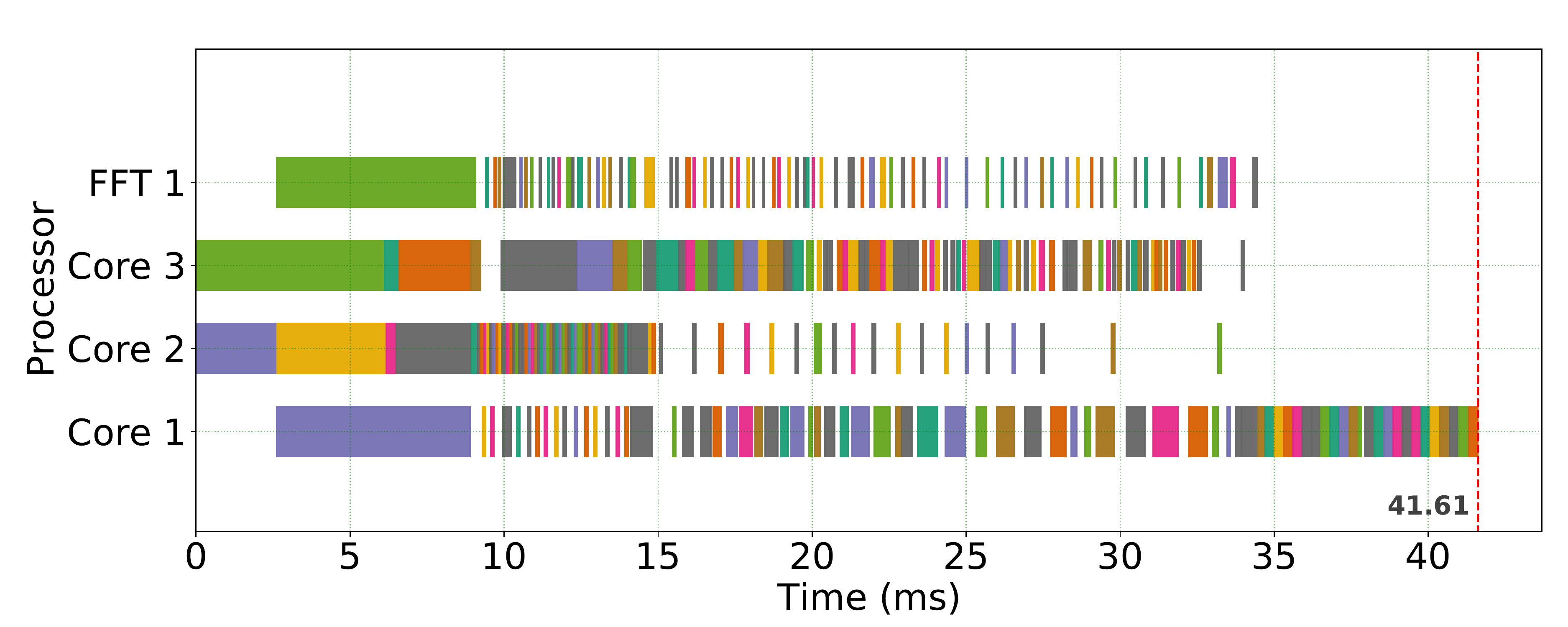}
         \vspace{-3mm}
         \caption{RoundRobin - RR}
         \label{fig:parallel_radar_correlators_rr}
     \end{subfigure}
     % \hfill
     \begin{subfigure}[b]{0.32\textwidth}
         \centering
         \includegraphics[width=0.95\columnwidth]{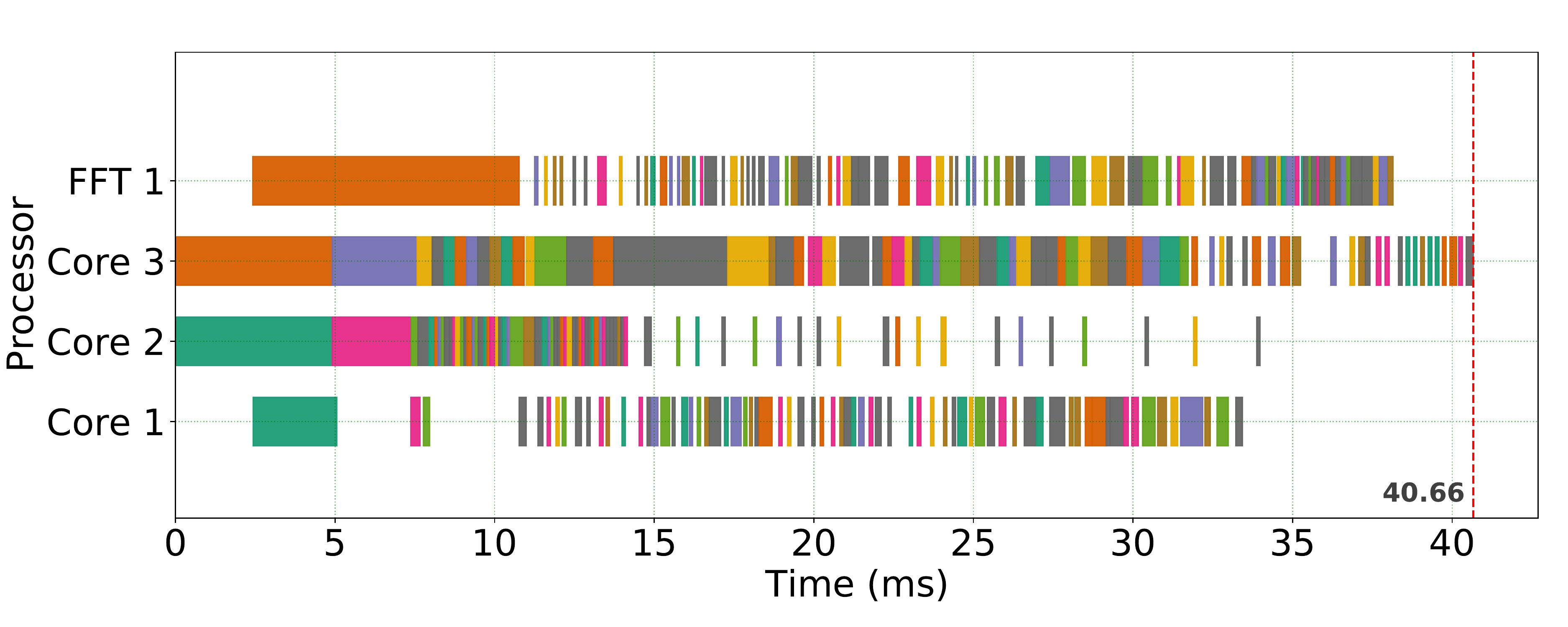}
         \vspace{-3mm}
         \caption{Earliest Finish Time - EFT}
         \label{fig:parallel_radar_correlators_eft}
     \end{subfigure}
     \caption{100 instances of auto-parallelized Radar Correlator application running on ZCU102 FPGA using three types of schedulers. Coloring encodes the application instance number modulo 10.}
\vspace{-6mm}
     \label{fig:parallel_radar_correlators}
\end{figure*}

% \begin{figure}
%     \centering
%     \includegraphics[width=\linewidth]{images/RadarCorr_and_WiFi_EFT_exp58.pdf}
%     \vspace{-6mm}
%     % \caption{Simultaneous execution of automatically parallelized Radar Correlator with serial WiFi TX application}
%     \caption{\josh{Demonstration of inter and intra-application parallelism via simultaneous execution of auto-parallelized Radar Correlator with serial WiFi TX}}
%     \label{fig:radar_correlator_and_wifi_tx_eft}
% \vspace{-6mm}
% \end{figure}
\begin{figure}
    \centering
      \begin{subfigure}[b]{0.5\textwidth}
         \centering
          \includegraphics[width=0.9\columnwidth]{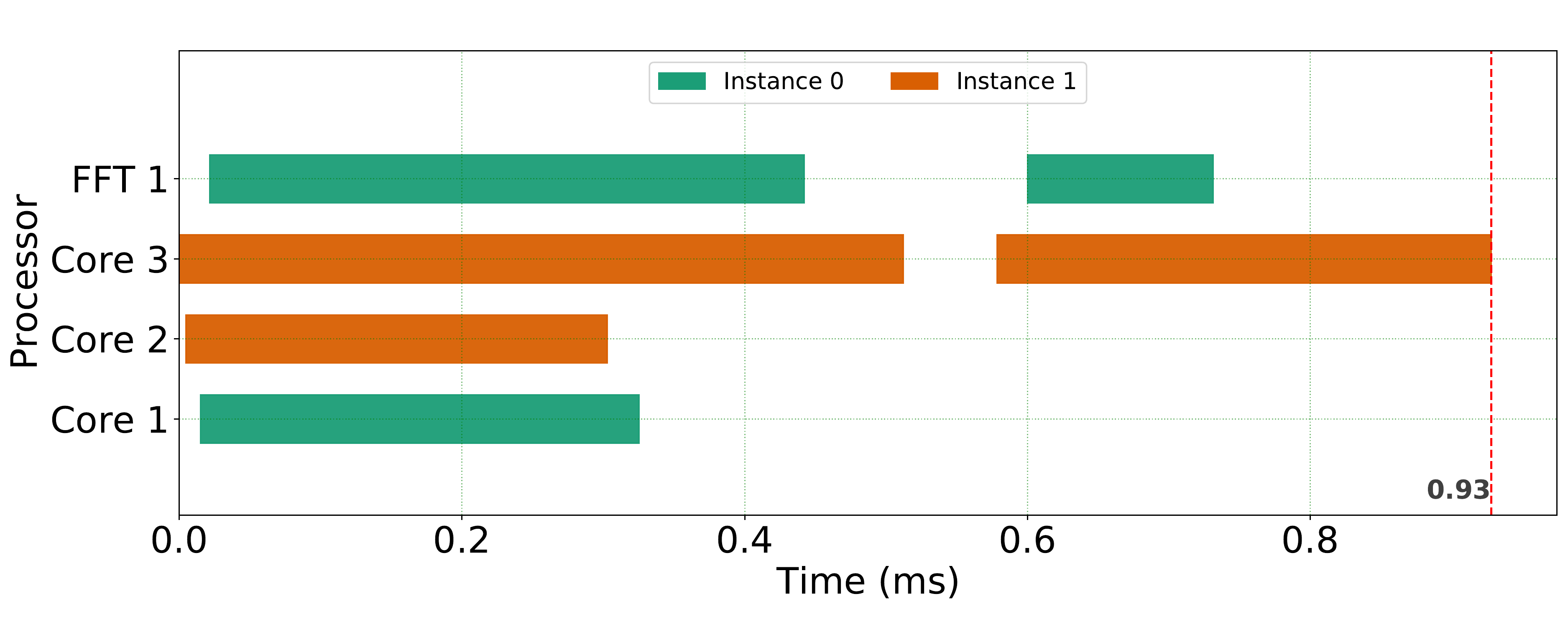}
         \vspace{-3mm}
         \caption{Two instances of parallel Radar Correlator running with EFT scheduler.}
         \label{fig:two_parallel_radar_correlator_eft}
     \end{subfigure}   
     \hfill
     \begin{subfigure}[b]{0.5\textwidth}
         \centering
        \includegraphics[width=0.9\columnwidth]{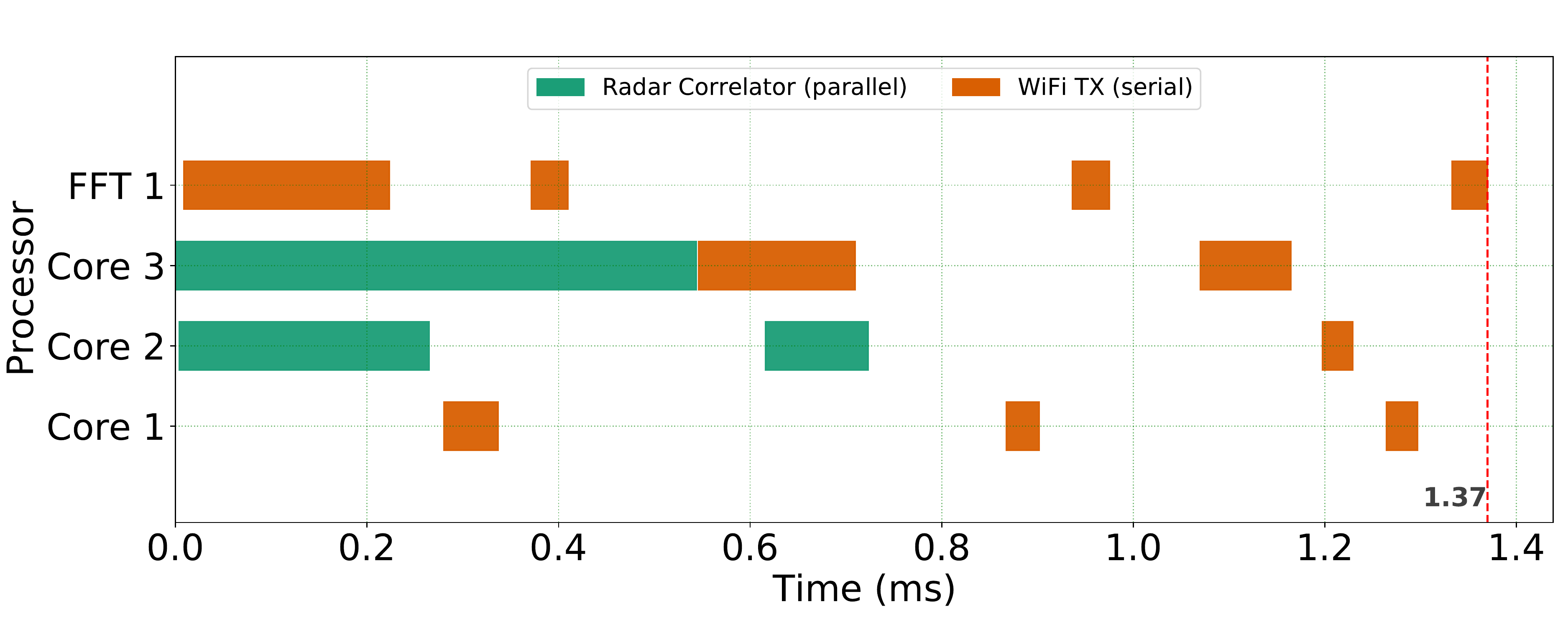}
        \vspace{-3mm}
         \caption{ Single instances of auto-parallelized Radar Correlator with serial WiFi TX.}
         \label{fig:radar_correlator_and_wifi_tx_eft}
     \end{subfigure}
    \caption{ Demonstration of inter and intra-application parallelism via simultaneous execution  on ZCU102 FPGA}
\vspace{-4mm}
    \label{fig:parallel_radar_correlator}
\end{figure}

\subsection{Functional Verification and Performance Analysis}
Pulse Doppler shows higher degree of parallelism relative to the other applications we use in this study. Therefore, we start with functional verification based on the Pulse Doppler execution through DS3 based simulation over an SoC with 4 and 8 FFT accelerators as shown in Figure~\ref{fig:4-pulse-4fft} and Figure~\ref{fig:4-pulse-8fft}, respectively. Since the 4-pulse version has 8 FFTs during the first stage (Figure~\ref{fig:4-pulse-4fft}), an SoC with  4 FFT accelerators takes  two rounds to complete the first stage, whereas on the 8 FFT configuration it takes one round. These two figures together illustrate ability to launch FFT tasks in parallel. Figure~\ref{fig:256-pulse-ds3} shows the 256-Pulse Doppler execution over the 8 FFT SoC configuration where first stage of the execution takes 64 rounds to complete and subsequent two phases take 32 rounds each. 
Overall, the plots in Figure~\ref{fig:pulse-gantt} show that our tool flow is able to extract parallelism in each stage of the execution and distribute the parallel tasks over the available compute resources. 
%\liang{The first experiment aims to validate the parallelization results using the Pulse Doppler application whose DAG is shown in %Figure~\ref{fig:DAG-figure}. 
%Figure~\ref{fig:4-pulse-gantt} gives the Gantt chart results when the 4-Pulse Doppler is mapped to SoCs with one processor core and 4 and 8 FFT accelerators respectively. All results are generated by DS3. We see that the FFTs on each stage of the DAG are distributively mapped to the different FFT accelerators.}

We demonstrate the execution of the real 256-Pulse Doppler implementation that is compiled for execution over the 8-core Xeon processor configuration using Figure~\ref{fig:256-pulse-cpu}, where execution of the parallelized application is managed by the CEDR. This plot shows same flow as Figure~\ref{fig:256-pulse-ds3} with three stage execution where first stage takes 64 rounds and subsequent two stages take 32 rounds each. This plot validates the runtime's ability to distribute FFT tasks as expected. Here we note that the DS3 based execution shows faster execution time than the Xeon processor based execution. There are two key factors to this observation. First the DS3 uses a model based execution where the FFT compute time is based on its actual execution over the Xilinx FFT IP Core synthesized for the ZCU102 FPGA. The compute time for the FFT accelerator is 128 ns whereas the execution time for the FFT task on the Xeon processor is 25,000 ns. Furthermore, overhead associated with the CEDR environment in terms of parsing application binary and dispatching task to the compute resources contribute to 
the increased overall execution time. Such runtime overhead is not modeled in DS3.

\begin{table}[t]
\centering
\begin{tabular}{|l|cc|cc|}
\hline
\multicolumn{1}{|c|}{\multirow{2}{*}{Application}} & \multicolumn{2}{c|}{Whole Application} & \multicolumn{2}{c|}{Only Tasks}     \\ \cline{2-5} 
\multicolumn{1}{|c|}{}                             & \multicolumn{1}{c|}{Xeon CPU}   & DS3  & \multicolumn{1}{c|}{Xeon CPU} & DS3 \\ \hline
Radar Correlator    & \multicolumn{1}{c|}{9.6}  & 14.8 & \multicolumn{1}{c|}{30.5} & 91.6 \\ \hline
Pulse Doppler       & \multicolumn{1}{c|}{59.0} & 67.3 & \multicolumn{1}{c|}{73.5} & 84.0 \\ \hline
Temporal Mitigation & \multicolumn{1}{c|}{3.9}  & 15.5 & \multicolumn{1}{c|}{48.8} & 97.5 \\ \hline
WiFi-TX             & \multicolumn{1}{c|}{17.0} & 21.2 & \multicolumn{1}{c|}{23.4} & 29.4 \\ \hline
Average             & \multicolumn{1}{c|}{22.4} & 29.7 & \multicolumn{1}{c|}{44.1} & 75.6 \\ \hline
\end{tabular}
\caption{ Execution time reduction (\%) with respect to the serial single Xeon core-based execution covering the end-to-end execution of the whole application and the time spent only over the FFT or GEMM tasks.} 
\vspace{-4mm}
\label{tab:speed-result}
\end{table}

Finally, Table~\ref{tab:speed-result}  shows reduction in execution time with respect to the serial execution for each application through CEDR on Xeon processor and DS3 in terms of time taken by the whole application and time spent only for the Type-1 and Type-2 tasks excluding data loads/stores from/to the disk. Consistent with the observations on the Pulse Doppler plots (Figure~\ref{fig:256-pulse-ds3} and Figure~\ref{fig:256-pulse-cpu}), as shown in Table~\ref{tab:speed-result},  for the computation tasks, we see an average of 44\% and 75.6\%  reduction in execution time with respect to the serial execution on the x86-based multi-core processor and  DS3-based simulation respectively. For the whole application,  the average reduction in execution time drops to 22.4\% and 29.7\% respectively, because of the reading from the disk and writing to the disk that contribute to the prolonged execution time. Note that WiFi-TX has less time saving compared to other applications. The reason is that WiFi-TX time spent on FFT is relatively less as this application involves other compute phases such as Scrambler, Interleaver, and Pilot-Insertion.

\subsection{FPGA-based Emulation and Analysis}

In Figure~\ref{fig:parallel_radar_correlators}, we show 
timing analysis for the Radar Correlator that is refactored into parallel execution form through our compiler tool chain.
The three plots illustrate the makespan for completing a workload composed of 100 instances of Radar Correlator executed based on the MET, RR, and EFT schedulers. This experiment shows that making a greedy choice favoring the accelerator (common in hand-crafted code) only results in FFT tasks starving for the accelerator to become available. MET enforces serial execution on the accelerator and also serves as a baseline for total execution time. The RR and EFT schedulers allow for utilizing the parallelism embedded into the application binary, and each one completes the workload in 41.61ms (1.48x faster) and 40.66ms (1.51x faster), respectively.

Figure~\ref{fig:parallel_radar_correlator} illustrates the runtime system's ability to execute FFT tasks from two applications as well as the independent FFTs within an application concurrently. In Figure~\ref{fig:two_parallel_radar_correlator_eft}, we show execution for two instances of Radar Correlator arriving concurrently where a total of four FFT tasks are dispatched to all PEs concurrently.
%We attribute the latency difference between single  and two Radar Correlator to the sensitivity of total execution time to the increased overhead in scheduling time with the use of EFT over the RR scheduler for this lightweight workload.
Finally in Figure~\ref{fig:radar_correlator_and_wifi_tx_eft}, we show the PE utilization based on the concurrent execution of WiFi TX and Radar Correlator applications. The EFT in this case favors the FFT accelerator to be used by the WiFi TX as it has higher latency than the Radar Correlator. It splits the use of ARM cores among the two applications and assigns Cores 2 and 3 to WiFi TX after the completion of the Radar Correlator.

\section{Conclusions}
In order to make heterogeneous SoCs accessible, there is the need for integrating the application development, compilation and runtime  processes vertically towards a unified ecosystem that enables productive application deployment without requiring users to become hardware experts in the process. Towards this goal, in this study, we present an integrated flow where we expose parallelism in a user application through dynamic profiling and memory analysis, and design a flexible binary structure where an application task can be invoked on any of its supported processing elements in the target SoC. 
We pass the parallelized binary to the runtime system that handles parsing dynamically arriving applications, scheduling tasks, and completing the workloads. We validate our approach through real-life radar applications executed on a diverse set of platforms. We believe that our integrated end-to-end system allows hardware-agnostic application development, enables exposing parallelism automatically in the user application and successful deployment on both multi-core homogeneous and heterogeneous architectures. 
The proposed dynamic analysis based method is sensitive to the nature of the inputs that may trigger different control flow for the same application. Future works will focus on addressing this open problem by merging the \emph{Control DAG} for different inputs and setting up a back-up control flow to recover the program execution and resolve unpredictable application behavior.

\label{sec:conclusions}
\bibliographystyle{IEEEtran}
\bibliography{IEEEabrv,ref}

\begin{thebibliography}{10}
\providecommand{\url}[1]{#1}
\csname url@samestyle\endcsname
\providecommand{\newblock}{\relax}
\providecommand{\bibinfo}[2]{#2}
\providecommand{\BIBentrySTDinterwordspacing}{\spaceskip=0pt\relax}
\providecommand{\BIBentryALTinterwordstretchfactor}{4}
\providecommand{\BIBentryALTinterwordspacing}{\spaceskip=\fontdimen2\font plus
\BIBentryALTinterwordstretchfactor\fontdimen3\font minus
  \fontdimen4\font\relax}
\providecommand{\BIBforeignlanguage}[2]{{%
\expandafter\ifx\csname l@#1\endcsname\relax
\typeout{** WARNING: IEEEtran.bst: No hyphenation pattern has been}%
\typeout{** loaded for the language `#1'. Using the pattern for}%
\typeout{** the default language instead.}%
\else
\language=\csname l@#1\endcsname
\fi
#2}}
\providecommand{\BIBdecl}{\relax}
\BIBdecl

\bibitem{abadi2016tensorflow}
M.~Abadi, P.~Barham, J.~Chen, Z.~Chen, A.~Davis, J.~Dean, M.~Devin,
  S.~Ghemawat, G.~Irving, M.~Isard \emph{et~al.}, ``Tensorflow: A system for
  large scale machine learning,'' in \emph{12th USENIX symposium on operating
  systems design and implementation (OSDI 16)}, 2016, pp. 265--283.

\bibitem{ragan2013halide}
J.~Ragan-Kelley, C.~Barnes, A.~Adams, S.~Paris, F.~Durand, and S.~Amarasinghe,
  ``Halide: a language and compiler for optimizing parallelism, locality, and
  recomputation in image processing pipelines,'' \emph{Acm Sigplan Notices},
  vol.~48, no.~6, pp. 519--530, 2013.

\bibitem{kotsifakou2018hpvm}
M.~Kotsifakou, P.~Srivastava, M.~D. Sinclair, R.~Komuravelli, V.~Adve, and
  S.~Adve, ``Hpvm: Heterogeneous parallel virtual machine,'' in
  \emph{Proceedings of the 23rd ACM SIGPLAN Symposium on Principles and
  Practice of Parallel Programming}, 2018, pp. 68--80.

\bibitem{chi2021extending}
Y.~Chi, L.~Guo, J.~Lau, Y.-k. Choi, J.~Wang, and J.~Cong, ``Extending
  high-level synthesis for task-parallel programs,'' in \emph{2021 IEEE 29th
  Annual International Symposium on Field-Programmable Custom Computing
  Machines (FCCM)}.\hskip 1em plus 0.5em minus 0.4em\relax IEEE, 2021, pp.
  204--213.

\bibitem{ketterlin2012profiling}
A.~Ketterlin and P.~Clauss, ``Profiling data-dependence to assist
  parallelization: Framework, scope, and optimization,'' in \emph{2012 45th
  Annual IEEE/ACM International Symposium on Microarchitecture}.\hskip 1em plus
  0.5em minus 0.4em\relax IEEE, 2012, pp. 437--448.

\bibitem{kim2010sd3}
M.~Kim, H.~Kim, and C.-K. Luk, ``Sd3: A scalable approach to dynamic
  data-dependence profiling,'' in \emph{2010 43rd IEEE/ACM International
  Symposium on Microarchitecture}.\hskip 1em plus 0.5em minus 0.4em\relax IEEE,
  2010, pp. 535--546.

\bibitem{wang2014integrating}
Z.~Wang, G.~Tournavitis, B.~Franke, and M.~F. O'boyle, ``Integrating
  profile-driven parallelism detection and machine-learning-based mapping,''
  \emph{ACM Transactions on Architecture and Code Optimization (TACO)},
  vol.~11, no.~1, pp. 1--26, 2014.

\bibitem{garcia2011kremlin}
S.~Garcia, D.~Jeon, C.~M. Louie, and M.~B. Taylor, ``Kremlin: Rethinking and
  rebooting gprof for the multicore age,'' \emph{ACM SIGPLAN Notices}, vol.~46,
  no.~6, pp. 458--469, 2011.

\bibitem{dagum1998openmp}
L.~Dagum and R.~Menon, ``Openmp: an industry standard api for shared-memory
  programming,'' \emph{IEEE computational science and engineering}, vol.~5,
  no.~1, pp. 46--55, 1998.

\bibitem{butenhof1997programming}
D.~R. Butenhof, \emph{Programming with POSIX threads}.\hskip 1em plus 0.5em
  minus 0.4em\relax Addison-Wesley Professional, 1997.

\bibitem{arda2020ds3}
S.~E. Arda, A.~Krishnakumar, A.~A. Goksoy, N.~Kumbhare, J.~Mack, A.~L. Sartor,
  A.~Akoglu, R.~Marculescu, and U.~Y. Ogras, ``Ds3: A system-level
  domain-specific system-on-chip simulation framework,'' \emph{IEEE
  Transactions on Computers}, vol.~69, no.~8, pp. 1248--1262, 2020.

\bibitem{uhrie2020automated}
R.~Uhrie, C.~Chakrabarti, and J.~Brunhaver, ``Automated parallel kernel
  extraction from dynamic application traces,'' \emph{arXiv preprint
  arXiv:2001.09995}, 2020.

\bibitem{MackTECS22}
\BIBentryALTinterwordspacing
J.~Mack, , S.~Hassan, N.~Kumbhare, M.~Gonzales, and A.~Akoglu, ``{CEDR} - {A}
  {Compiler-integrated}, {Extensible} {DSSoC} {Runtime},'' \emph{ACM
  Transactions on Embedded Computing Systems}, Apr. 2022. [Online]. Available:
  \url{https://doi.org/10.1145/3529257}
\BIBentrySTDinterwordspacing

\bibitem{lattner2004llvm}
C.~Lattner and V.~Adve, ``Llvm: A compilation framework for lifelong program
  analysis \& transformation,'' in \emph{International Symposium on Code
  Generation and Optimization,}.\hskip 1em plus 0.5em minus 0.4em\relax IEEE,
  2004, pp. 75--86.

\bibitem{aladdinShao2014}
Y.~S. Shao, B.~Reagen, G.-Y. Wei, and D.~Brooks, ``Aladdin: A pre-rtl,
  power-performance accelerator simulator enabling large design space
  exploration of customized architectures,'' in \emph{Proceeding of the 41st
  Annual International Symposium on Computer Architecuture}, ser. ISCA
  '14.\hskip 1em plus 0.5em minus 0.4em\relax IEEE Press, 2014, p. 97–108.

\bibitem{Wasabi2019Lehmann}
D.~Lehmann and M.~Pradel, ``Wasabi: A framework for dynamically analyzing
  webassembly,'' in \emph{Proceedings of the Twenty-Fourth International
  Conference on Architectural Support for Programming Languages and Operating
  Systems}, ser. ASPLOS '19.\hskip 1em plus 0.5em minus 0.4em\relax New York,
  NY, USA: Association for Computing Machinery, 2019, p. 1045–1058.

\bibitem{luk2005pin}
C.-K. Luk, R.~Cohn, R.~Muth, H.~Patil, A.~Klauser, G.~Lowney, S.~Wallace, V.~J.
  Reddi, and K.~Hazelwood, ``Pin: building customized program analysis tools
  with dynamic instrumentation,'' \emph{Acm sigplan notices}, vol.~40, no.~6,
  pp. 190--200, 2005.

\bibitem{augonnet2011starpu}
C.~Augonnet, S.~Thibault, R.~Namyst, and P.-A. Wacrenier, ``{StarPU}: a unified
  platform for task scheduling on heterogeneous multicore architectures,''
  \emph{Concurrency and Computation: Practice and Experience}, vol.~23, no.~2,
  pp. 187--198, 2011.

\bibitem{donyanavard_sparta_2016}
B.~Donyanavard, T.~Mück, S.~Sarma, and N.~Dutt, ``Sparta: Runtime task
  allocation for energy efficient heterogeneous manycores,'' in \emph{2016
  International Conference on Hardware/Software Codesign and System Synthesis
  (CODES+ISSS)}, 2016, pp. 1--10.

\bibitem{donyanavard_sosa_2019}
B.~Donyanavard, T.~M\"{u}ck, A.~M. Rahmani, N.~Dutt, A.~Sadighi, F.~Maurer, and
  A.~Herkersdorf, ``Sosa: Self-optimizing learning with self-adaptive control
  for hierarchical system-on-chip management,'' in \emph{Proceedings of the
  52nd Annual IEEE/ACM International Symposium on Microarchitecture}, ser.
  MICRO '52.\hskip 1em plus 0.5em minus 0.4em\relax New York, NY, USA:
  Association for Computing Machinery, 2019, p. 685–698.

\bibitem{maity_2021_SEAMSSelfOptimizing}
B.~Maity, B.~Donyanavard, A.~Surhonne, A.~Rahmani, A.~Herkersdorf, and N.~Dutt,
  ``{SEAMS}: {Self}-{Optimizing} {Runtime} {Manager} for {Approximate} {Memory}
  {Hierarchies},'' \emph{ACM Transactions on Embedded Computing Systems},
  vol.~20, no.~5, pp. 48:1--48:26, Jul. 2021.

\bibitem{martins_hierarchical_2019}
A.~L. del Mestre~Martins, A.~H.~L. {da Silva}, A.~M. Rahmani, N.~Dutt, and
  F.~G. Moraes, ``Hierarchical adaptive multi-objective resource management for
  many-core systems,'' \emph{Journal of Systems Architecture}, vol.~97, pp.
  416--427, 2019.

\bibitem{tan_picos++_2019}
X.~Tan, J.~Bosch, C.~Álvarez, D.~Jiménez-González, E.~Ayguadé, and
  M.~Valero, ``A hardware runtime for task-based programming models,''
  \emph{IEEE Transactions on Parallel and Distributed Systems}, vol.~30, no.~9,
  pp. 1932--1946, 2019.

\bibitem{moazzemi_2019_HESSLEFREEHeterogeneous}
K.~Moazzemi, B.~Maity, S.~Yi, A.~M. Rahmani, and N.~Dutt, ``{HESSLE}-{FREE}:
  Heterogeneous systems leveraging fuzzy control for runtime resource
  management,'' \emph{ACM Transactions on Embedded Computing Systems}, vol.~18,
  no.~5s, pp. 74:1--74:19, Oct. 2019.

\end{thebibliography}
\end{document}